\documentclass[12pt]{iopart}
\usepackage{graphicx,stackengine,cite}
\usepackage[colorlinks,linkcolor=blue,citecolor=blue]{hyperref}  \usepackage{url}
\usepackage{verbatim}
\usepackage{xcolor}
\usepackage{bbold}
\usepackage[normalem]{ulem}

\newcommand{\bQ}{\bi{\mathcal{Q}}}
\newcommand{\Q}{\mathcal{Q}}

\newcommand{\vi}{\bi{v}}

\newcommand{\x}{\bi{x}}
\newcommand{\y}{\bi{y}}

\newcommand{\ri}{\bi{r}}
\newcommand{\rh}{\hat{\bi{r}}}
\newcommand{\eS}{\bi{S}}
\newcommand{\q}{\bi{q}}

\newcommand{\vg}{\bgamma}

\newcommand{\lt}{\left}
\newcommand{\rt}{\right}

\newcommand{\key}{\bi{k}}

\newcommand{\Fr}{\mathcal{F}}

\newcommand{\eps}{\boldsymbol{\epsilon}}

\newcommand{\vr}{\bi{r}}
\newcommand{\p}{\bi{p}}

\newcommand{\n}{\hat{\bi{n}}}

\newcommand{\calV}{{\cal V}}
\newcommand{\calF}{{\cal F}}
\newcommand{\calFbulk}{{\cal F}_{\rm bulk}}
\newcommand{\calFcore}{{\cal F}_{\rm core}}

\newcommand{\Fbulk}{{F}_{\rm bulk}}
\newcommand{\Fcore}{{F}_{\rm core}}

\newcommand{\barbphi}{ \boldsymbol{\bar{\phi}} }

\newcommand{\Correction}[1]{{\color{black}#1}}

\usepackage{iopams}
\usepackage{setstack} 
\usepackage{tocloft} 
\usepackage{subcaption}

\begin{document}

\title[Dynamical theory of topological defects II: Universal aspects of defect motion]{Dynamical theory of topological defects II: Universal aspects of defect motion}

\author{Jacopo Romano$^1$, Beno\^it Mahault$^{1,*}$, Ramin Golestanian$^{1,2,\dagger}$}

\address{$^1$Max Planck Institute for Dynamics and Self-Organization (MPI-DS), 37077 G\"ottingen, Germany}
\address{$^2$Rudolf Peierls Centre for Theoretical Physics, University of Oxford, Oxford OX1 3PU, United Kingdom}
\ead{$^{*}$benoit.mahault@ds.mpg.de}
\ead{$^{\dagger}$ramin.golestanian@ds.mpg.de}

\vspace{10pt}

\begin{indented}
\item[] \today
\end{indented}

\begin{abstract}
We study the dynamics of topological defects in continuum theories governed by a free energy minimization principle, building on our recently developed framework [Romano J, Mahault B and Golestanian R 2023 J. Stat. Mech.: Theory Exp. 083211]. We show how the equation of motion of point defects, domain walls, disclination lines and any other singularity can be understood with one unifying mathematical framework. For disclination lines, this also allows us to study the interplay between the internal line tension and the interaction with other lines. This interplay is non-trivial, allowing defect loops to expand, instead of contracting, due to external interaction. We also use this framework to obtain an analytical description of two long-lasting problems in point defect motion, namely the scale dependence of the defect mobility and the role of elastic anisotropy in the motion of defects in liquid crystals. For the former, we show that \Correction{the effective defect mobility} is strongly problem-dependent, but it can be computed with high accuracy for a pair of annihilating defects. For the latter, we show that at the first order in perturbation theory, anisotropy causes a non-radial force, making the trajectory of annihilating defects deviate from a straight line. At higher orders, it also induces a correction in the mobility, which becomes non-isotropic for the $+1/2$ defect. We argue that, due to its generality, our method can help to shed light on the motion of singularities in many different systems, including driven and active non-equilibrium theories.
\end{abstract}

\pagebreak

\tableofcontents

\title[Dynamical theory of topological defects II: Universal aspects of defect motion]{}

\begin{center}
   \rule{10cm}{1pt} 
\end{center}

\setcounter{page}{1}

\section{Introduction}

Topological defects are singular configurations appearing in symmetry-broken phases \cite{Nelson2002defects}, ranging from trapped quantum gases \cite{HadzibabicNature2006} to cosmic scales~\cite{Hindmarsh_1995}. They are important in condensed matter physics, as exemplified in the key roles they play in coarsening dynamics~\cite{Bray2002review}, two dimensional melting~\cite{Kosterlitz1973JPC}, and magnetic properties of type-II superconductors~\cite{Abrikosov2004RMP}.
Furthermore, topological defects feature in various phenomena in active matter \cite{uchida2010,Wensink2012,SanchezNature2012,Giomi2013,Thampi2013} and biology ~\cite{Kawaguchi2017nature,Saw2017Nature,MaroudasSacks2020,CopenhagenNatPhys2021}.

In strongly ordered systems, defects are usually described over macroscopic scales as quasi-particles in interaction with the surrounding order parameter phase field~\cite{PismenBook}.
The derivation of this reduced description from the dynamics of the full order parameter has been the subject of many studies~\cite{Dafermos1970,IMURA1973403,Eshelby1980PhilMagA,AmbegaokarPRB1980,Dubois-violette,Kawasaki1983linedef,Kawasaki1984Progr,KAWASAKI1984319,BODENSCHATZ1988PhysD,NEU1990PhysD,Pismen1990PRA,Rubinstein1991,RodriguezPRA1991,Pismen1991PhysicaD,DennistonPRB1996,Pleiner1988PRA,Semenov_1999EPL,NajafiEPJB2003,Radzihovsky2015PRL},
while the ubiquity and dynamical characteristics of defects in active matter have led to a recent revival of interest in this problem~\cite{PismenPRE2013,TangSM2019,CortesePRE2018,ShankarPRL2018,Vafadefects2020,ZhangPRE2020,AnghelutaNJP2021,VafaSoftMatt2022}.
One difficulty in carrying out such coarse-graining procedure is that defects are intrinsically microscopic structures, 
such that their description {\it a priori} requires the knowledge of the order parameter dynamics over microscopic scales. 
Hence, many studies have relied on matched asymptotics by solving the field theory in the vicinity and far away from the defect, 
while the continuity of the full solution is imposed in an intermediate matching region (for a pedagogical introduction, see Ref.~\cite{PismenBook}).
On the other hand, when the defect core size is made truly microscopic, there is no guarantee that it is faithfully described by the phenomenological theories expressed in terms of smoothly varying fields.

The effective equations of motion governing the dynamics of defects generally take a similar functional form~\cite{PismenBook}, suggesting a certain degree of universality insofar as the details of the microscopic core structure only set the value of certain coefficients in the reduced description.
Existing results are, however, mostly restricted to idealized cases (except, e.g., Ref. \cite{Gartland2002ws}), and thus omit important features present in real systems.
For example, while most approaches consider the limit of slow defects, we have shown that significant memory effects emerge due to the dependencies of the order parameter landscape on the past position and velocity of the defect~\cite{romano2023}.
Another feature of liquid crystals generally ignored is the effect of elastic anisotropy, which introduces higher order nonlinearities to the order parameter field theory.
Anisotropy in the elastic response of the medium, on the other hand, is responsible for qualitative changes in the dynamics of defects~\cite{Sven2003PRL,Brugues2008PRL,MissaouiPRR2020,Review_LC_Harth2020} that cannot be accounted for in the single Frank constant approximation. 

Here, we propose a new approach allowing to derive the dynamical equations of motion for defects from any dissipative field theory that satisfies a minimization principle. Importantly, this approach is formally valid at all orders in the defect velocity and for any free energy functional describing the dynamics.
In Sec.~\ref{sec_exact_var}, we demonstrate that, under a set of rather weak assumptions, the defect equation of motion takes a universal form as the details of the core structure set the value of a unique length scale in the expression of the mobility,
while the effective force moving the defect is fully determined by the large scale physics. 

To illustrate the power of the approach, we apply it to multiple scenarios involving either point or line defects. We show in particular that the main features of the physics of topological defects can be captured by means of a low mobility expansion, 
which leads to substantial simplifications in the equations of motion. Whereas most of existing results correspond to the leading order contribution to this expansion, we show in Sec.~\ref{sec_def_GL} how in simple cases improved approximations can be obtained by considering higher order corrections. Applying the method to a theory describing nematic liquid crystals in two dimensions, we moreover quantify in Sec.~\ref{sec_def_NL} how elastic anisotropy spontaneously rotates pairs of annihilating defects and affects their mobilities. Section~\ref{sec_lines} is finally devoted to defect lines. A derivation of the Allen-Cahn equation~\cite{ALLEN1979} for domain walls is firstly given in Sec.~\ref{sec_dw}, while the dynamics of disclination lines and loops emerging in three-dimensional phases with broken $U(1)$ symmetry are discussed in Sec.~\ref{sec_discl_lines}. 

\section{Dynamics of point topological defects in two dimensions}
\label{sec_top_def}
\subsection{Derivation of the defect dynamics from free energy variations}
\label{sec_exact_var}

Throughout this section, we study a two-dimensional system described by an order parameter $\bphi(\x,t)$ whose dynamics minimizes a free energy $\calF = \int \rmd^2\x \, F(\bphi,\nabla\bphi)$
as described by 
\begin{equation} \label{eq_EL}
    \partial_t \bphi=-\frac{\delta \Fr}{\delta \bphi}.
\end{equation}
\Correction{Equation~\eref{eq_EL} corresponds to a deterministic version of \textit{model A} in the Halperin-Hohenberg classification~\cite{HH1977}, and thus serves as a general form to describe any relaxational dynamics without conservation law. 
For example, in dynamics with broken polar or nematic orientational order, $\bphi$ corresponds to a vectorial or rank-2 tensor field:
\begin{equation*}
    \bphi_{\rm pol} = \rho
    \left(\begin{array}{c} 
    \cos\theta \\ \sin\theta
    \end{array}\right), \qquad
    \bphi_{\rm nem} = \frac{\rho}{\sqrt{2}}
    \left(\begin{array}{c c} 
    \cos2\theta & \sin2\theta \\ 
    \sin2\theta & -\cos2\theta
    \end{array}\right),
\end{equation*}
where $\rho(\x,t)$ and $\theta(\x,t)$ respectively set the magnitude and orientation of order.
On the other hand, phases with broken $\mathbb{Z}_2$ symmetry will be described by a scalar order parameter.   


In what follows, we work in a parameter regime for which the system is strongly ordered far away from the defects.
In practice, this implies that the free energy $\calF$ reduces to a known functional $\calFbulk$ that depends only on the slow modes of the dynamics.
The free energy $\calFbulk$ then describes the dynamics of $\bphi$ everywhere except in the vicinity of defects, where it is captured by a {\it a priori} unknown free energy $\calFcore$.
For instance, as will be detailed further in Sec.~\ref{sec_def_GL}, for systems with orientational order the strongly ordered limit corresponds to the case where the norm $\rho$ of $\bphi$ is a fast mode, while $\calFbulk$ depends only on the orientation $\theta$ and its derivatives.

In the remaining of this section, we focus on the case of point defects such as those occurring in two-dimensional phases with orientational order, while the case with discrete symmetry will be addressed in Sec.~\ref{sec_lines}.
We consider a configuration with an arbitrary number of defects, and derive the equation of motion of a specific defect whose position and velocity are denoted respectively as $\q(t)$ and $\vi(t)$. For convenience, we will use a generalized scalar product simply defined as the sum of squared components: $|\bphi|^2 = \bphi \cdot \bphi \equiv \phi_b \phi_b$.
Throughout this work, summation over repeated indices is implied.}

\begin{figure}[t!] 
    \centering
    \includegraphics[width=.6\linewidth]{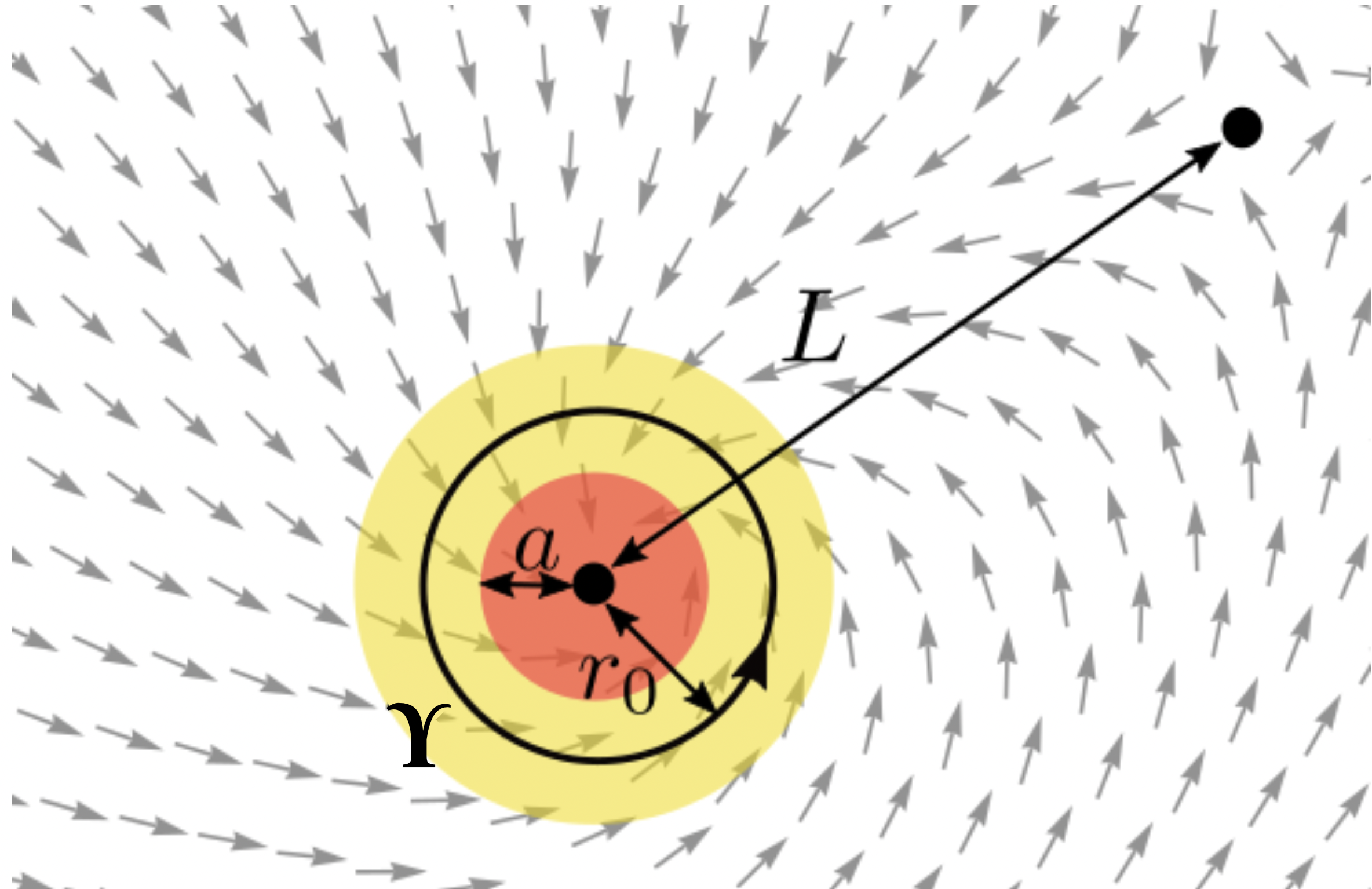}
    \caption{
   Schematics of the three scales $a$, $r_0$, and $L$ involved in the derivation. The microscopic scale $a$ sets the defect core size (red region).
   The macroscopic scale $L$ is given by the system size, or the typical distance between defects. 
   In the intermediate matching region parametrized by $r_0$ (yellow), it is assumed that both the core and bulk theories hold.}
\label{sketch_scales}
\end{figure}

Below, we present a detailed derivation of the equation of motion for defects which applies to an arbitrary free energy $\calF$ satisfying the following assumptions:
\begin{enumerate}
\item \label{ass:1} {\it Translational invariance:} the free energy density $\Fcore$ varies in space only through the field $\bphi$, so that it is translationally invariant.
\item \label{ass:2} {\it Microscopic core size:} the dynamics of the core is associated with a length scale $a \to 0$, which plays the role of the core size. 
As $|\bphi|$ varies from zero at the centre of the core to one over a distance $a$, gradients of $\bphi$ in the core are of order $a^{-1}$.
\item \label{ass:3} {\it The rigid core assumption:} at the leading order in $a$, the shape of the core is independent of its position or velocity. Up to ${\cal O}(a)$ terms, $\bphi(\x,t)$ can thus be expressed in the reference frame of the core in terms of a fixed function $\barbphi(\y)$ with $\y = \bi{R}^{-1}(t)[\x - \q(t)]/a$ and where the rotation matrix $\bi{R}(t)$ parametrizes the direction of the core. The relationship between $\bphi$ and $\barbphi$ depends on the nature of the order. 
For vectorial and nematic (rank-2 tensor) orders, we respectively have $\bphi_{\rm pol}=\bi{R}\barbphi_{\rm pol}$ and $\bphi_{\rm nem}=\bi{R}^{-1}\barbphi_{\rm nem}\bi{R}$.
Note that, regardless of the type of order, $|\bphi| =  |\barbphi|$.
\item \label{ass:4} {\it Existence of a matching regime:} lastly, we assume the existence of an intermediate region at distance $\sim r_0$ from the core where $\calFbulk$ and $\calFcore$ 
coincide and both describe the dynamics. This region shall be well-separated from the other scales of the problem, such that $a \ll r_0 \ll L$ where $L$ stands for a macroscopic scale, e.g.\ the system size or the typical inter-defect distance (see a sketch in Fig.~\ref{sketch_scales}).
\end{enumerate}

Assumption~\eref{ass:1} is natural so long as the system is not externally driven by a spatially dependent field\footnote{(\ref{ass:1}) should still be verified in the presence of weak or smoothly varying fields. 
Formally, we require that the external field varies on length scales $\gg a$ and timescales $\gg a^2/D$, where $D$ is the effective diffusivity associated with the dynamics of $\bphi$.},
while~\eref{ass:2} ensures a proper scale separation between the defect core size and the macroscopic dynamics. 
\eref{ass:4} reflects the continuity condition between the core and bulk physics, and plays a central role in the matched asymptotic methods~\cite{PismenBook}.
Assumption~\eref{ass:3}, in fact, follows from~\eref{ass:1} and~\eref{ass:2}. 
Indeed, for vanishing $a$ the field $\bphi$ at the core can be generally expanded in powers of $a/L$, with $L$ a problem-dependent macroscopic scale. 
The leading order of this expansion is by construction independent of the relative positions and velocities of other defects, as they contribute to terms at least ${\cal O}(a/L)$.
Moreover, from~\eref{ass:1} the order parameter at the core is, up to rotations and translations, uniquely determined by $\Fcore$, which eventually leads to~\eref{ass:3}.

\subsubsection{The force applied on a defect core}
\label{sec_force_def}
To find the equation of motion for the defect core, we first express the variational equation~\eref{eq_EL} in an integral form as
\begin{equation}
\label{var1}
     \int \rmd^2\x \; \delta \bphi \cdot \partial_t \bphi=-\delta \calF ,
\end{equation}
where $\delta \bphi$ is a fixed boundary condition perturbation.
By construction, Eq.~\eref{var1} is valid for any infinitesimal $\delta \bphi$. 
Here, we consider a specific type of perturbation:
\begin{equation} \label{pert_phi}
\Correction{\delta \bphi(\x,t) = f(\x,t) (\delta\q \cdot \nabla) \bphi(\x,t),}
\end{equation}
where $\delta \q$ is an infinitesimal vector and $f(\x,t)$ is a smooth interpolating envelope function equal to one at \Correction{$\x = \q(t)$}, which decays quickly to zero for \Correction{$|\x - \q(t)| > r_0$},
such that it is zero at the boundaries of the system and at the positions of other defects. 
The introduction of the envelope function $f$ is done to isolate the core for which we wish to derive the equation of motion.
It is not strictly necessary, but simplifies the derivation by allowing us to discard the effect of the variation $\delta \bphi$ at the system boundaries and at the other defect cores.
Discarding this prefactor, would in fact lead to additional ${\cal O}(a/L)$ contributions to the final equation, which are subdominant in the limit of well separated scales.

Defining a closed curve $\Upsilon$ within the matching region, we split the l.h.s.\ and r.h.s.\ of Eq.~\eref{var1} into contributions from inside and outside $\Upsilon$,
corresponding respectively to the core and the bulk.
Namely,
\begin{equation}
\label{var2}
    \int_{\rm core}\rmd^2 \x \, \delta \bphi \cdot \partial_t \bphi + \int_{\rm bulk}\rmd^2\x\, \delta \bphi \cdot \partial_t \bphi =
    - \delta \calFcore -\delta \calFbulk.
\end{equation}
We now re-express the bulk free energy variation as
\begin{equation} \label{var2bis}
\delta\calFbulk = \int_{\rm bulk}\rmd^2\x \, \delta \bphi \cdot \frac{\delta\calFbulk}{\delta \bphi} - \oint_{\Upsilon} \rmd S_i \, \delta\bphi \cdot \frac{\partial\Fbulk}{\partial (\partial_{i} \bphi)}, 
\end{equation}
where the second term on the r.h.s.\ is a surface contribution retained after integrating by parts, and where we assumed that the free energy density 
$\Fbulk(\bphi,\nabla\bphi)$ does not depend on higher derivatives of $\bphi$\footnote{The generalization to cases where $\Fbulk$ depends on higher derivatives of $\bphi$ is straightforward and does not affect the final equation~\eref{var3}.}. 
As $\delta\bphi$ is zero outside the core region, this surface term only includes a contribution from the boundary $\Upsilon$, 
while the preceding minus sign implies that $\rmd\eS$ points to the outside of the core.

To compute $\delta\calFcore$, we note that in the core $\delta\bphi = (\delta\q \cdot \nabla)\bphi$ as by construction $f(|\x-\q| \le r_0) = 1$.
Therefore, $\delta\bphi$ in the core corresponds to an infinitesimal translation.
Since, from~\eref{ass:1}, $\Fcore$ is translationally invariant, it follows that 
\begin{equation*}
\delta\Fcore = \frac{\partial \Fcore}{\partial \phi_b}(\delta\q\cdot\nabla)\phi_b
+\frac{\partial \Fcore}{\partial (\partial_{j}\phi_b)}(\delta\q\cdot\nabla)\partial_{j}\phi_b + ... = (\delta\q\cdot\nabla) \Fcore ,
\end{equation*}
where the dots stand for dependencies of $\Fcore$ {on} higher order derivatives of $\bphi$, if any.
Hence, $\delta\calFcore$ is an exact differential, and therefore, we obtain
\begin{equation} \label{var2bisbis}
    \delta \calFcore = \oint_{\Upsilon} \rmd\eS \cdot \delta \q \, \Fcore = \oint_{\Upsilon} \rmd\eS \cdot \delta \q \, \Fbulk,
\end{equation}
where the second equality results from the fact that $\Upsilon$ belongs to the matching region.
Putting together Eqs.~(\ref{var2},\ref{var2bis},\ref{var2bisbis}), we find that
\begin{equation}
    \label{var3}
     \int_{\rm core}\rmd^2\x \,\partial_t \bphi \cdot \delta \bphi = \delta q_j \oint_{\Upsilon}\rmd {S}_i \, {T}_{{\rm bulk},ij} ,
\end{equation}
where we have defined the canonical stress tensor of the bulk theory \Correction{~\cite{toupin1960stress,susskind2017special}}:
\begin{equation*} \label{eq_def_Tbulk}
    {T}_{{\rm bulk},ij} \equiv  \frac{\partial\Fbulk}{\partial (\partial_{i} \bphi)}\cdot (\partial_{j}\bphi) - \delta_{ij} \Fbulk.
\end{equation*}
Hence, the r.h.s.\ of Eq.~\eref{var3} corresponds to the net momentum flux  through the matching region, 
and is solely determined by the large-scale bulk physics.

\subsubsection{The defect friction tensor} As we show now, the l.h.s.\ of~\eref{var3} weakly depends on the specific form of the core free energy.
We first note that Eq.~\eref{var3} holds for an arbitrary choice of curve $\Upsilon$ in the matching region.
Choosing without loss of generality $\Upsilon$ to be a circle of radius $r_0$ around the singularity, we thus expect that the final result will be independent of $r_0$.
Using the rigid core assumption~\eref{ass:3}, it is clear that in the core $\partial_t \bphi = -(\vi \cdot \nabla)\bphi$\footnote{The time derivative of $\bphi$ also includes a contribution from the rotation matrix $\bi{R}(t)$. However, this contribution is subdominant in the limit $r_0/a \gg 1$.}, such that the l.h.s.\ of Eq.~\eref{var3} is given by
\begin{equation}
\label{mob}
\int_{\rm core}\rmd^2\x \,\partial_t \bphi \cdot \delta \bphi
= - v_i \delta q_j R_{ik}(t)R_{jl}(t)\int_{\rm {\cal D}_{r_0/a}}\rmd^2\y \, \partial_{k}\barbphi\left(\y\right) \cdot \partial_{l}\barbphi\left(\y\right) ,
\end{equation}
where we have used the change of variable 
$\y = \bi{R}^{-1}(t)(\x - \q(t))/a$ 
and ${\cal D}_{r_0/a}$ is the disk of radius $r_0/a$ centred at $\bf 0$.
Equation~\eref{mob} defines the effective \Correction{friction tensor} of the defect dynamics:
\begin{equation} \label{def_mu}
\fl\qquad \zeta_{ij}\left(\frac{r_0}{a},t\right) \equiv R_{ik}(t)R_{jl}(t)\int_{{\cal D}_{r_0/a}}\rmd^2\y \, \partial_{k}\barbphi\left(\y\right) \cdot \partial_{l}\barbphi\left(\y\right)
\equiv R_{ik}(t)R_{jl}(t) \bar{\zeta}_{kl}\left(\frac{r_0}{a}\right).
\end{equation}
We note that, since $\barbphi$ is fully determined by the core structure, it is independent of the macroscopic defect variables such at its position, velocity and orientation.
The tensor $\bar{\bzeta}$ is thus a fixed function which specifies the structure of the defect friction, and can be evaluated at leading order in $a$ from the static single defect solution.
Namely, differentiating~\eref{def_mu} w.r.t.\ $r_0$ and parametrizing $\y$ with the polar coordinates $(r_0,\varphi)$, we find that
\begin{equation} \label{barmushape1}
\frac{\rmd \bar{\zeta}_{ij}}{\rmd r_0} = r_0 \int_0^{2\pi} \rmd\varphi \, \left. \partial_{i}\barbphi\left(r_0,\varphi\right) \cdot \partial_{j}\barbphi\left(r_0,\varphi\right)\right|_{\rm ssd},
\end{equation}
where the `ssd' subscript indicates that the integrand is calculated from the {\it static single defect} solution of the bulk theory, since the right hand side of this equation is evaluated in the matching region.
As already noted in a number of works~\cite{Pismen1990PRA,Bray2002review}, the functional shape of the \Correction{ friction tensor} is only determined by the bulk theory, while the core theory only enters through an integration constant when~\eref{barmushape1} is integrated on both sides.
This integration constant plays the role of a phenomenological parameter that captures the microscopic features of the core.

The time dependency of $\bzeta$ results from the fact that an anisotropic core structure may lead the defect to experience different friction strengths in different directions. 
Although $\bzeta$ is determined by the {\it a priori} unknown core free energy $\calFcore$, we show in~\ref{mobilandsymm} that in most relevant situations $\bzeta$ is often isotropic and thus independent of the defect orientation (see, however, Sec.~\ref{sec_mob_alpha2} for a counter example). In the following, we therefore keep the time dependency of $\bzeta$ implicit.

\subsubsection{The general equation of motion for defects} 
Gathering the results accumulated so far, and noting that Eq.~\eref{var3} must be valid for all $\delta\q$,
we find that the defect equation of motion takes the compact form
\begin{equation}
    \label{var5}
    \bzeta\left(\frac{r_0}{a}\right) \vi = - \oint_{{\cal C}_{r_0}}\bi{T}^T_{{\rm bulk}} \, \rmd \eS,
\end{equation}
where ${\cal C}_{r_0}$ stands for the circle of radius $r_0$ centred at the defect core.
Equation~\eref{var5} highlights that the defect equation of motion, up to a constant factor in the mobility, 
is universal as it does not depend on the core physics so long as the latter satisfies translational and rotational invariance.
This equation moreover bears a transparent physical interpretation, since it simply states that the momentum flux through the boundary of the core
is, up to frictional effects, entirely dissipated into the motion of the defect, which is a natural consequence of~\eref{ass:3}.
The r.h.s.\ of Eq.~\eref{var5} corresponds to the Ericksen force defined by Eshelby~\cite{Eshelby1980PhilMagA}, 
which moreover takes the same formal expression as the Peach-Koehler force acting on dislocations~\cite{Eshelby1980PhilMagA,LUBARDA20191550}.
Lastly, it is important to note that both sides of Eq.~\eref{var5} depend on the matching variable $r_0$, while the actual equation of motion of the defect must be independent of it, since $r_0$ is arbitrary.
In fact, we show below that eliminating $r_0$ self-consistently allows in some cases to fix the functional form of the mobility.

We now illustrate how~\eref{var5} can be used to explicitly derive the dynamics of defects. We start by showing how standard results are recovered for systems described by the archetypal Ginzburg-Landau free energy~\eref{genfreen}. We then address nonlinear problems such as when defects dynamics evolve in a medium featuring elastic anisotropy. 

\subsection{Defects in the classical Ginzburg-Landau framework}\label{sec_def_GL}

\subsubsection{Elimination of the matching scale} \label{sec_def_GL_trajs}

The simplest choice for $\calF$ is certainly the Ginzburg-Landau free energy
\begin{equation}
\label{genfreen}
    \calF^{\rm GL} = \int \rmd^2 \x \, \left[ \frac{1}{2}|\nabla \bphi|^2 + \chi(1-|\bphi|^2)^2 \right] \, .
\end{equation}
Note that we work in time units such that the coefficient in front of the elastic term in~\eref{genfreen} has been set to one.
The scale of the defect core is then given by $a \simeq \chi^{-1/2}$.
Over scales much larger than $a$, the dynamics of the order parameter is then fully captured by that of its orientation $\theta$, which is associated with the bulk free energy
\begin{equation}
\label{freeb_iso}
    \calFbulk^{\rm GL} = \frac{1}{2}\int \rmd^2\x\, |\nabla \theta|^2 .
\end{equation}
The dynamics of $\theta$ is thus ruled by the diffusion equation, while
the stress tensor associated with~\eref{freeb_iso} is given by
$T_{{\rm bulk},ij}^{\rm GL}(\theta)=(\partial_{i}\theta)(\partial_{j}\theta)-\frac{1}{2}\delta_{ij}|\nabla \theta|^2$.
In particular, we showed previously~\cite{romano2023} that the orientation field gradient induced by a defect of charge $s$ following a trajectory $\q(t)$ with velocity $\vi(t)$ is given by
\begin{equation}
    \label{eq_gradthetaGL}
    \nabla\theta(\x,t)=-\frac{s}{2}\eps \int_{-\infty}^{t} \frac{\rmd t' }{(t-t')} \left[\nabla+\vi(t')\right]e^{-\frac{|\x-\q(t')|^2}{4(t-t')}},
\end{equation}
where $\bepsilon$ denotes the two-dimensional antisymmetric Levi–Civita tensor
\begin{equation*}
    \bepsilon = \left(\begin{array}{cc}
         0 & -1 \\
         1 & 0
    \end{array}\right).
\end{equation*}
Using the linearity property of the diffusion equation, the total orientation field landscape induced multiple defects corresponds to the linear superposition of single defect solutions~\eref{eq_gradthetaGL}.

Since the matching region is assumed to be well separated from the other scales of the theory, we now explicitly evaluate both sides of Eq.~\eref{var5} in the limit of $a \to 0$ and $r_0 \to 0$, keeping $r_0 \gg a$.
To obtain the r.h.s., we note that the presence of a defect leads to a divergence of $|\nabla \theta| \sim a^{-1}$ at the core, such that for $r_0$ small the contour integral is dominated by the discontinuous part of $\nabla \theta$.
Although the general expression~\eref{eq_gradthetaGL} is nonlocal in time, the discontinuous part $\nabla\theta_{\rm d}$ of $\nabla\theta$ in the vicinity of a defect
is formally determined by its instantaneous position and velocity~\cite{romano2023}, namely
\begin{equation}
\label{discpart}
    \nabla\theta_{\rm d}(\ri,t) \underset{r\to 0}{\simeq} s\eps\left[\frac{ \rh}{r}+\frac{\vi(t)}{2}\ln\lt(\frac{r}{\lambda(t)}\rt)-\frac{1}{2}\left(\vi(t)\cdot\rh\right)\rh\right],
\end{equation}
where we have defined $\ri \equiv \x - \q(t)$, $r\equiv |\ri|$ and $\rh \equiv \ri/r$.
The additional length scale $\lambda(t)$ leads to a continuous contribution to~\eref{discpart}, but was included for dimensional consistency.
This quantity formally depends on the whole knowledge of the past history of the defect trajectory, and it is generally not possible to evaluate it directly from~\eref{eq_gradthetaGL}.
For now, we thus retain it as a phenomenological constant, and we will show in the following sections how it can be determined or approximated.

Denoting $\nabla\theta=\nabla\theta_{\rm d}+\nabla\theta_{\rm c}$, with $\nabla\theta_{\rm c}$ accounting for the continuous part of the gradient, we calculate in~\ref{isotropicforce} the r.h.s.\ of Eq.~\eref{var5}, which leads for $r_0\to 0$ to
\begin{equation} \label{eq_iso_r0}
     \left[\zeta_{ij}\left(\frac{r_0}{a}\right)
     + \pi s^2 \ln\lt(\frac{e^{\frac{1}{2}}\lambda(t)}{r_0}\rt) \delta_{ij}
     \right]v_j =  
    - 2\pi s\epsilon_{ij} \partial_j \theta_{\rm c}(\q(t),t) .
\end{equation}
It is clear that only the terms on the l.h.s.\ of Eq.~\eref{eq_iso_r0} depend on the matching variable $r_0$. 
Hence, we conclude that the friction $\zeta_{ij} = \zeta \delta_{ij}$ while the quantity $r_0 \exp[-\zeta/(\pi s^2)]$ must remain independent of $r_0$.
Furthermore, from its definition~\eref{def_mu} the defect friction is independent of any macroscopic scale, which yields
\begin{equation*}
    \zeta\left( \frac{r_0}{a} \right) = \pi s^2 \ln\left( \frac{r_0}{\bar{a}} \right),
\end{equation*}
where $\bar{a} \propto a$ is a phenomenological constant that plays the role of the core effective radius, and depends on the details of the core physics.
It is straightforward to show that this expression of $\zeta$ satisfies~\eref{barmushape1}, 
while the same result could have been obtained directly by calculating the integral in~\eref{barmushape1} using the solution~\eref{discpart} with $\vi = \bf{0}$.
The equation of motion for the defect therefore reads
\begin{equation}
\label{isomotion}
    \ln\lt(\frac{e^{\frac{1}{2}}\lambda(t)}{\bar a}\rt)\vi(t) = -\frac{2}{s}\, \eps\, \nabla \theta_{\rm c}(\q(t),t).
\end{equation}

Equation~\eref{isomotion} is formally similar to Eq.~\eref{var5}.
However, whereas the latter depends on the arbitrary matching variable $r_0$ through $\bzeta$ and the integration contour on the r.h.s.,
Eq.~\eref{isomotion} determines the defect motion independently of the choice of matching region.
The only quantity that depends on the core physics in~\eref{isomotion} is the parameter $\bar{a}$. 
Its value is set by the core free energy $\calFcore$, which sets the profile of the full order parameter $\bphi$ at the core.
For the Ginzburg-Landau free energy~\eref{genfreen}, it has been shown that $\bar{a} \sqrt{\chi}\simeq 1.126$~\cite{PismenBook}.
The coefficient $\lambda(t)$ in the expression of the effective friction, on the other hand, is a macroscopic scale fixed by the history of the defect trajectory~\cite{romano2023}.
The r.h.s.\ of Eq.~\eref{isomotion} shows that defects are essentially moved by gradients of the orientation field~\cite{Eshelby1980PhilMagA}.
This gradient is in general generated by other defects, or by specific anchoring conditions at the system boundaries. 
Its expression and that of $\lambda(t)$ can in principle be obtained by solving for the dynamics of the orientation field $\theta$ with the appropriate boundary conditions at the defect cores.
Below, we show that the terms of Eq.~\eref{isomotion} can be explicitly determined in particular configurations.

\subsubsection{An isolated defect moving at constant velocity}
\label{sec_def_const_vel}
We first study the simple case of a defect moving uniformly with velocity $\vi = \bar{\vi}$.
An experimental realization of this case would for example consist of a defect subject to an imposed spin wave $\theta_{\rm ext}(\x) \equiv \key \cdot \x$, with $\key$ the corresponding wavevector,
induced by an external field. 
Under these conditions, the total angular field is given by the sum of $\theta_{\rm ext}(\x)$ and the uniformly moving defect solution to the diffusion equation, namely~\cite{DennistonPRB1996,Radzihovsky2015PRL,romano2023}
\begin{equation}
\label{constvel}
    \nabla\theta(\vr)=\frac{s}{2} e^{-\frac{1}{2} \bar{\vi}\cdot\vr} \eps \left[\bar{\vi} K_0\left(\frac{\bar{v}r}{2}\right)-\bar{v}\hat\vr K_1\left(\frac{\bar{v}r}{2}\right)\right]+\key,
\end{equation}
where $K_0$ and $K_1$ are modified Bessel functions of the second kind while, as previously, $s$ denotes the charge of the defect and $\vr = \x-\q(t)$.
Expanding Eq.~\eref{constvel} for $r \to 0$ and keeping only nonvanishing terms, we find that $\nabla\theta_{\rm c} = \key$ while the discontinuous part $\nabla\theta_{\rm d}$ is given by~\eref{discpart} with $\lambda = 4 \bar{v}^{-1} e^{-\gamma_{\rm E}}$; and where $\gamma_{\rm E} \approx 0.578$ denotes the Euler–Mascheroni constant. 

Replacing these expressions in Eq.~\eref{isomotion}, we thus recover the classical result~\cite{PismenBook}
\begin{equation} \label{const_speed_eom}
    \ln\left(4\frac{e^{\frac{1}{2}-\gamma_{\rm E}}}{\bar{v}\bar{a}}\right)\bar{\vi} = 
    -\frac{2}{s}\,\eps \,\key,
\end{equation}
whose solution determines the velocity of the defect in response to an imposed spin wave. Consistently with the general result~\eref{isomotion} and as noted by a number of authors~\cite{IMURA1973403,BODENSCHATZ1988PhysD,Pismen1990PRA,Ryskin1991PRL,DennistonPRB1996,Bray2002review,Brugues2008PRL,PismenPRE2013,Radzihovsky2015PRL,TangSM2019}, the defect is subject to nonlinear friction. As we show below, this feature is rather generic, while the exact expression of the friction depends on the problem of interest.

\subsubsection{Self-consistent solution for a pair of slowly moving defects}
\label{sec_sc_two_defects}

We now consider a pair of defects with opposite charges $s_\pm = \pm s$.
To proceed further, we note from~\eref{isomotion} that the defect mobility $\mu \equiv \zeta^{-1} \sim \ln^{-1}(\lambda/a)$.
Hence, so long as the macroscopic scale $\lambda$ remains much larger than the core radius, defects move relatively slowly with a speed $v = {\cal O}(\mu)$.
In what follows, we thus treat $\mu$ as a small parameter, which allows us to derive an approximate expression for $\lambda$. 

At first order in $\mu$, the angular field created by a moving defect takes the universal form given by~\eref{discpart}, regardless of its trajectory~\cite{romano2023}. 
The only trajectory-dependent quantity is then the parameter $\lambda(t)$.
Denoting, respectively, $\q_\pm(t)$ and $\vi_\pm(t)$ as the positions and velocities of the two defects, it then follows from~\eref{isomotion} that
\begin{equation} \label{eq_lowmob_2def}
    \fl \qquad\qquad \ln\left(\frac{e^{\frac{1}{2}}\lambda_\pm}{\bar{a}}\right)\vi_\pm 
    = -\frac{2}{s_\pm}\eps\nabla\theta_{\mp} (\q_\pm(t),t) = \mp\frac{\hat{\q}}{q} +  \ln\left( \frac{\lambda_\mp}{2q} \right)\vi_\mp  + \left(\vi_\mp\cdot\hat{\q}\right)\hat{\q} ,
\end{equation}
where $\theta_{\pm}$ denote the orientation field landscapes generated by the positively and negatively charged defects, 
while the second equality was obtained by replacing $\theta_\pm$ with~\eref{discpart}.
$\lambda_\pm$ in Eq.~\eref{eq_lowmob_2def} denote the scales associated with the $\pm$ defects, while we have also defined $\q \equiv \frac{1}{2}(\q_+ - \q_-)$.
It is a straightforward exercise to show that~\eref{eq_lowmob_2def} implies that $\vi_+ = -\vi_- = v \hat{\q}$, such that the speed of both defects follows
\begin{equation} \label{eq_veff_2def}
    v = -\frac{\mu_{\rm eff}}{q}, 
    \qquad \mu_{\rm eff} \equiv 
    \left[\ln\left( \frac{e^{\frac{3}{2}}\lambda_+\lambda_-}{2\bar{a}q } \right)\right]^{-1}.
\end{equation}
As $q(t)$ is the only macroscopic scale of the problem, we propose the ansatz: $\lambda_\pm(t) \propto q(t)$. 
Hence, it follows that $\dot{\mu}_{\rm eff} \sim v \mu_{\rm eff}^2 \sim \mu_{\rm eff}^3$ such that, at first order in the mobility, the angular field generated by a defect moving according to~\eref{eq_veff_2def} can be calculated from~\eref{eq_gradthetaGL} treating $\mu_{\rm eff}$ as a constant parameter. 
The details of this calculation are presented in Ref.~\cite{romano2023}, while the resulting expression for $\nabla\theta$ is again formally given by~\eref{discpart} with 
$\lambda_+(t) = \lambda_-(t) = 2\sqrt{2}q(t) \mu_{\rm eff}^{-\frac{1}{2}}e^{-\frac{1}{2}(1 + \gamma_{\rm E})}$.
Combining this result with the expression of $\mu_{\rm eff}$ given in~\eref{eq_veff_2def} yields the following self-consistent condition 
\begin{equation}\label{eq_cond_mu_2def}
\mu_{\rm eff}e^{\mu_{\rm eff}^{-1}}
    = \frac{4 q}{\bar{a}}e^{\frac{1}{2} - \gamma_{\rm E}} 
\end{equation}
whose solution can be expressed in terms of the velocity as $\mu_{\rm eff}^{-1} = \ln \left[4e^{\frac{1}{2}-\gamma_{\rm E}}/(v\bar{a})\right]$. 
As there is no other macroscopic scale but the defects' degrees of freedom, we recover an expression for the friction similar to that of~\eref{const_speed_eom}.
The condition~\eref{eq_cond_mu_2def}, however, only holds perturbatively in $\mu_{\rm eff} \, (\propto v)$.
Evaluating the solution of Eq.~\eref{eq_cond_mu_2def} up to terms ${\cal O}\left( \ln[\ln(q/\bar{a})]/\ln(q/\bar{a}) \right)$, we thus obtain
\begin{equation}
\label{consistentGLdynamics}
    \ln\left[ \frac{4 q}{\bar{a}}e^{\frac{1}{2}-\gamma_{\rm E}}\ln\left(\frac{4e^{\frac{1}{2}-\gamma_{\rm E}}}{\bar{a}} q\right)\right]v = -\frac{1}{q}.
\end{equation} 
Equation~\eref{consistentGLdynamics} fully determines the dynamics of the defect pair.

\begin{figure}[t!] 
    \centering\includegraphics[width=1.0\linewidth]{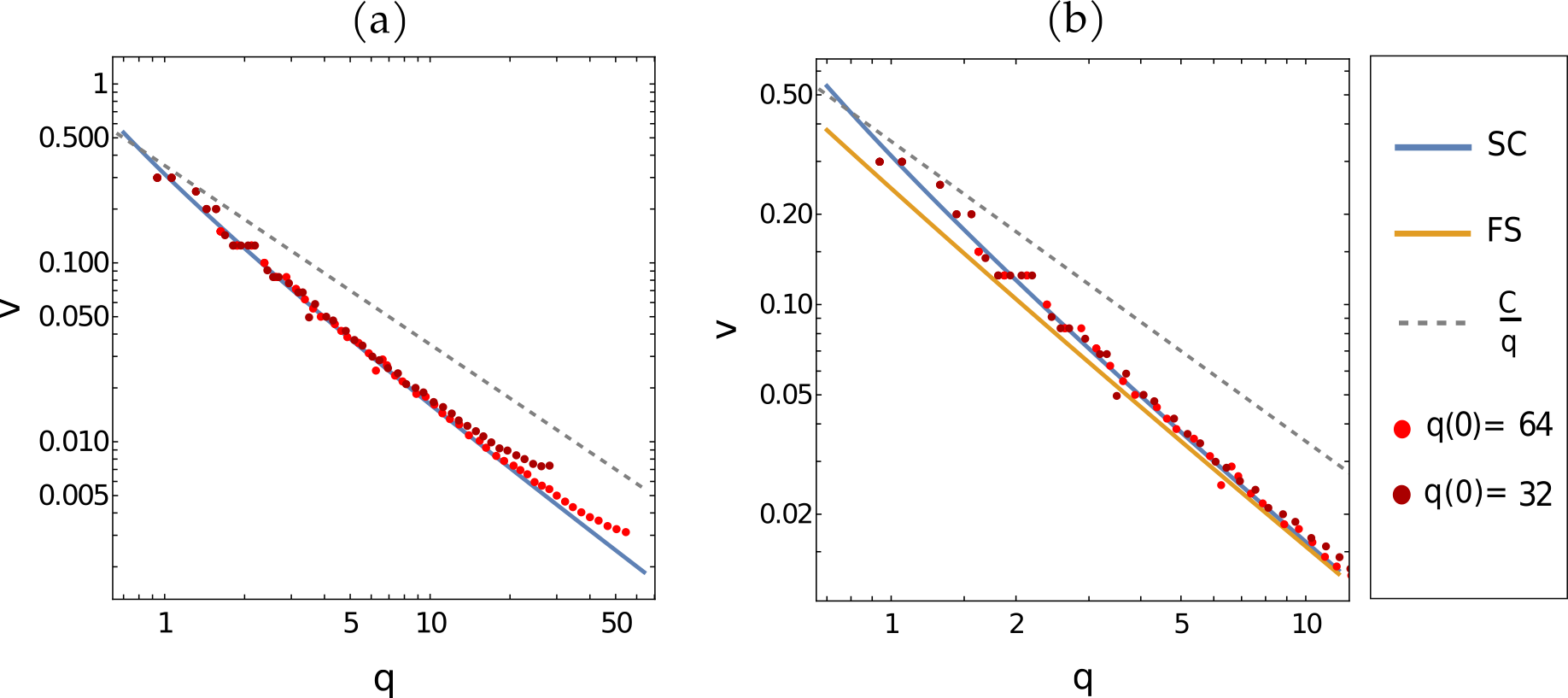}
    \caption{Comparison of trajectories between simulation (data points) and the self-consistent approximation~(\ref{consistentGLdynamics}) (solid blue line, SC) initial defect separation $q(0)=32$ and $q(0)=64$ ($\bar a \approx 0.36$). Panel
    (a): Defect velocity as function of its position over the full simulation range.
    Panel (b): Zoom of (a) closer to the annihilation point. The yellow line is obtained by replacing $q$ with a fixed scale (FS) $q=32$ in the nested logarithm of~(\ref{consistentGLdynamics}). This approximation differs from the numerical data close to annihilation, but improves at larger distances. 
    The dashed grey curves indicate the inverse distance scaling predicted by the Coulomb interaction.}
    \label{GL_trajecotries}
\end{figure}

Extensive discussions on scale- \cite{Bray2002review,Brugues2008PRL,TangSM2019} or velocity- \cite{IMURA1973403,BODENSCHATZ1988PhysD,Pismen1990PRA,Ryskin1991PRL,DennistonPRB1996,Brugues2008PRL,PismenPRE2013,Radzihovsky2015PRL,TangSM2019} dependent defect mobility can be found in the literature. The derivation outlined above shows that, in fact, the scale $\lambda$ entering the expression of the mobility is primarily determined by the relevant scales of the background orientation field. In the general case, moreover, $\lambda$ may depend on the past configurations of the system, leading to nontrivial memory effects~\cite{romano2023}.

We show in Figure~\ref{GL_trajecotries} comparisons between the relation $v(q)$ obtained from numerical simulations of a $s = \pm 1$ defect pair annihilation in a vectorial field described by the Ginzburg-Landau model (Eq.~\eref{genfreen}), and theoretical predictions with different approximation schemes for the defect mobilities (details on numerical simulations can be found in~\ref{app_numerics}). Since the simulations are initialized with the static defect profiles, the order parameter field first relaxes to the dynamical solution of~\eref{eq_EL}.
This effect induces a transient behaviour of $v(q)$ at large $q$ that depends on the initial defect separation.
Past this transient, $v(q)$ exhibits a behaviour independent of the initial defect separation, which we find in clear departure from the the $\propto q^{-1}$ scaling predicted by the constant mobility approximation (dashed lines in Fig.~\ref{GL_trajecotries}).
Conversely, a parameter-free comparison with Eq.~\eref{consistentGLdynamics} reveals excellent agreement with the measured relation $v(q)$ (blue curves in Fig.~\ref{GL_trajecotries}).

\subsubsection{Many defects and the collective mobility} 
In the low mobility approximation, Eq.~\eref{eq_lowmob_2def} can be generalized to an arbitrary number of defects as
\begin{equation} \label{eq_lowmob_Ndef}
s_\alpha^2 \ln\left(\frac{e^{\frac{1}{2}}\lambda_\alpha}{\bar{a}}\right)\vi_\alpha
    = \sum_{\beta \ne \alpha} s_\alpha s_\beta\left[ \frac{\hat{\q}_{\alpha \beta}}{q_{\alpha \beta}} - \ln\left( \frac{\lambda_\beta}{2q_{\alpha \beta}} \right)\vi_\beta  - \left(\vi_\beta\cdot\hat{\q}_{\alpha \beta}\right)\hat{\q}_{\alpha \beta} \right] ,
\end{equation}
where $\q_\alpha$, $\vi_\alpha$ and $s_\alpha$ denote the position, velocity and charge of the $\alpha$th defect, respectively, while $\q_{\alpha \beta} \equiv \frac{1}{2}(\q_\alpha - \q_\beta)$.
Rearranging the terms, we recast~\eref{eq_lowmob_Ndef} into a more compact form:
\begin{equation} \label{eq_lowmob_Ndef_compact}
	\vi_\alpha = \sum_\beta \bi{{\cal M}}_{\alpha \beta} \left(- \nabla_{\q_\beta} U_{\rm C}\right),
\end{equation}
where $U_{\rm C} \equiv -\sum_{\alpha < \beta} s_\alpha s_\beta \ln|q_{\alpha \beta}|$ is the two-dimensional Coulomb potential, 
while the mobility matrix $\bi{{\cal M}}$ is defined through $\sum_\gamma \bi{{\cal M}}_{\alpha \gamma} \bi{{\cal Z}}_{\gamma \beta} = \delta_{\alpha \beta}\bi{{\cal I}}$, where the friction matrix $\bi{{\cal Z}}$ is defined as 
\begin{equation*}
	\bi{{\cal Z}}_{\alpha \beta} \equiv s_\alpha s_\beta \left\{ \begin{array}{ll} \ln\left(\frac{e^{\frac{1}{2}}\lambda_\alpha}{\bar{a}}\right) \bi{{\cal I}} & \;\;\;\alpha = \beta \\ 
	\ln\left( \frac{\lambda_\beta}{2q_{\alpha \beta}} \right)\bi{{\cal I}} + \hat{\q}_{\alpha \beta}\hat{\q}_{\alpha \beta} &\;\; \;\alpha \ne \beta \end{array} \right. ,
\end{equation*}
and where $\bi{{\cal I}}$ denotes the two-dimensional identity matrix.

Equation~\eref{eq_lowmob_Ndef_compact} reveals that the many-body defect dynamics is coupled via the collective (non-diagonal) mobility $\bi{{\cal M}}$.
When $\lambda_\alpha \ne \lambda_\beta$, $\bi{{\cal M}}$ is moreover not symmetric, such that it introduces effective non-reciprocal couplings between the defect velocities.
Although the Coulomb force is conservative, the centre of mass of the system is thus generally not immobile. A similar effect was reported in the context of active nematics~\cite{Vafadefects2020}.
Here, we observe that this effect also arises in the absence of active drive, so that it is generic to the relaxation dynamics of systems with a large number of defects.
To rationalize this result, we note that both the interaction and the mobility in Eq.~\eref{eq_lowmob_Ndef_compact} are set by the orientation field landscape, which is itself driven out of equilibrium by the motion of the defects (see Eq.~\eref{eq_gradthetaGL}). During relaxation to equilibrium, the effective interactions between defects are therefore mediated by a nonequilibrium medium,
which is a well-identified mechanism for the generation of non-reciprocal couplings~\cite{Dzubiella2003PRL,soto2014PRL,Gupta2022PRE}.

Unlike the two-defect case, it is generally not possible to determine analytically the parameters $\lambda_\alpha(t)$ 
by calculating the exact solution for the angular field $\theta(\vr,t)$.
We however note that, since the $\lambda_\alpha$ are given by the macroscopic scales of the system (for example the mean inter-defect separation),
the mobility in~\eref{eq_lowmob_Ndef_compact} is dominated by its diagonal coefficients which are of order $\ln(\lambda/\bar{a})$,
while the off-diagonal coefficients are ${\cal O}(1)$.
At the leading order in $\ln(\lambda/\bar{a})$, one can thus replace $\lambda$ by {\it any} macroscopic scale of the problem.
Indeed, replacing $\lambda \to \lambda + \lambda'$ does not change the result to the leading order, namely
\begin{equation*}
	\ln\left( \frac{\lambda + \lambda'}{\bar{a}}\right) = \ln\left( \frac{\lambda}{\bar{a}} \right) + {\cal O}(1).
\end{equation*}
For two defects, Figure~\ref{GL_trajecotries}(b) shows that the approximation $\lambda \propto q$ matches well with the self-consistent solution~\eref{consistentGLdynamics} for large enough defect separation (compare the blue and yellow curves).
As we show in the following section, this approximation is moreover of great use when dealing with more complex problems.

\subsection{Defects in nonlinear systems}\label{sec_def_NL}

\subsubsection{The large $\ln(a)$ expansion}
\label{sec_loga_exp}

Although the terms of Eq.~\eref{var5} can be explicitly calculated in simple scenarios,
calculations quickly become intractable if the bulk free energy contains higher order nonlinearities.
Here, we thus show that Eq.~\eref{var5} admits a powerful expansion in the inverse of $\ln(a)$ related to the low mobility expansion discussed above.
We first note that, as the static single defect solution for the orientation field $\theta_{\rm ssd}(r,\varphi)$---where $(r,\varphi)$ stands for the polar coordinates in the defect frame---is scale-free by construction, we have $\theta_{\rm ssd}(r,\varphi) = \theta_{\rm ssd}(\varphi)$. Hence, Eq.~\eref{barmushape1} becomes
\begin{equation} \label{eq_barmu_der_2}
    \frac{\rmd \bar{\zeta}_{ij}}{\rmd r_0}
    =\frac{1}{r_0}\int_0^{2\pi} \rmd\varphi \, (\partial_{\varphi}\theta_{\rm ssd}(\varphi))^2
    \, \hat{\varphi}_{i}\hat{\varphi}_{j} \equiv \frac{\bar{\nu}_{ij}}{r_0},
\end{equation}
where $\hat{\bvarphi}$ is the unit vector tangent to the unit circle. Equation~\eref{eq_barmu_der_2} therefore implies that the friction matrix takes the form
\begin{equation} 
    \label{logmu}
    \bar{\zeta}_{ij}\left(\frac{r_0}{a}\right) = \bar{\nu}_{ij}\ln\left(\frac{r_0}{a}\right)+C_{ij},
\end{equation}
where both $\bar{\bnu}$ and $\bi{C}$ are dimensionless and independent of $a$. We are now in a good position to expand Eq.~\eref{var5} using $\ln(a)$ as a large parameter.
Such expansion---although it does not involve a ratio of two scales---shall be seen as a formal way to take the limit of slowly moving defects.
We note that at the leading order in $\ln(a)$, Eq.~\eref{logmu} simplifies to $\bar{\zeta}_{ij} = -\bar{\nu}_{ij}\ln(a) + {\cal O}(1)$, such that the \Correction{friction tensor} does no more involve the matching variable $r_0$ nor the constant $\bi{C}$.

Moreover, replacing ${\zeta}_{ij} \sim \ln(a)$ in Eq.~\eref{var5}, one observes that the defect velocity is at least of order $\ln^{-1}(a)$.
Hence, at the leading order the stress tensor on the r.h.s.\ of Eq.~\eref{var5} can be written as
${\bi{T}}_{\rm bulk} = \bi{T}_{\rm sb} + {\cal O}(\ln^{-1}(a))$, 
where the subscript `sb' indicates that $\bi{T}$ is calculated from the static solution of the bulk theory.
We also note that since the static solution satisfies $\delta \calFbulk/\delta\bphi_{\rm sb} = \mathbf{0}$ (from~\eref{eq_EL}), the corresponding stress tensor is divergence-free: $\nabla\cdot\bi{T}_{\rm sb} = \mathbf{0}$.
Consequently, the line integral on the r.h.s.\ of Eq.~\eref{var5} is independent of the chosen path, and we get
\begin{equation}
\label{eq_logaexp}
    \ln\left(\frac{\lambda}{a}\right)\nu_{ij}v_j = -\oint \rmd S_j \, T_{{\rm sb},ji},
\end{equation}
where $\nu_{ij}=R_{ik}(t)R_{jl}(t) \bar{\nu}_{kl}$ includes the possible effects of defect anisotropy (see Eq.~\eref{def_mu}) and the integral on the r.h.s.\ can be evaluated on any loop enclosing the defect core.
The scale $\lambda \gg a$ on the l.h.s.\ was included for dimensional consistency, and gives a subdominant contribution in $\ln(a)$.
At the leading order, $\lambda$ can thus formally be replaced by \emph{any} relevant macroscopic scale of the theory. 
We finally note that the $\ln(a)$ expansion provides a formulation that is independent of the matching variable $r_0$.
Contrary to~\eref{var5}, Eq.~\eref{eq_logaexp} can thus be directly used without the need to eliminate $r_0$ by enforcing the matching condition.

\subsubsection{Elastic anisotropy and misalignment-induced forces} 
\label{sec_misalignment_forces}

To illustrate how~\eref{eq_logaexp} can be employed to address more complex scenarios, we now consider the dynamics of a pair of topological defects in a nematic liquid crystal featuring elastic anisotropy.
The order parameter for this case is therefore the nematic tensor
\begin{equation*}
    \bQ(\rho,\theta) = \frac{\rho}{\sqrt{2}}\left(\begin{array}{cc}
\cos2\theta & \sin2\theta \\
\sin2\theta & -\cos2\theta  \\
\end{array}\right),
\end{equation*}
such that topological defects carry a half-integer charge: 
$s = \pm \frac{1}{2}$.
The bulk dynamics is described by the Frank-Oseen free energy~\cite{deGennes_Prost}, which takes the form\footnote{Equation~\eref{simplefrank} corresponds to the 
free energy density $F_{d} = \frac{1}{2}K_1(\nabla\cdot\hat{\bi{n}})^2
+ \frac{1}{2}K_2[\hat{\bi{n}} \times (\nabla\times\hat{\bi{n}})]^2$ 
expressed in terms of the director field $\hat{\bi{n}}$, where 
$\frac{1}{2}(K_1+K_2) =1$ and $(K_2-K_1)/(K_2+K_1) =\alpha$.}
\begin{equation}
\label{simplefrank}
    \calFbulk = \frac{1}{2} \int \rmd^2\x \,
    \left[ |\nabla \theta|^2 + \sqrt{2}\alpha \nabla \theta\cdot \bQ(1,\theta)\cdot \nabla \theta \right],
\end{equation}
where the parameter $\alpha \in [-1;1]$ is nonzero whenever 
splay and bend deformations incur different elastic costs.
Although elastic anisotropy commonly occurs in liquid crystals, the theoretical understanding of defect dynamics in this context remains limited~\cite{Sven2003PRL,MissaouiPRR2020}.
This feature results from the fact that, for $\alpha \ne 0$, the relaxational dynamics of $\theta$ follows a nonlinear equation,
such that deriving the counterpart of Eq.~\eref{eq_gradthetaGL} is generally difficult; even perturbatively in $\alpha$.
Here, we show how some progress can be made from Eq.~\eref{eq_logaexp} for slow defects and in the presence of weak anisotropy.

To calculate the defect mobility, we use~\eref{eq_barmu_der_2} and consider the static single defect solution associated with the free energy~\eref{simplefrank}. This solution can be expressed in an implicit form as~\cite{LandauToE}
\begin{equation}
    \label{implnem}
    \partial_\varphi\theta_{\rm ssd}(\varphi)= 1 + \frac{1}{p}
    \sqrt{ \frac{ 1-\alpha p^2\cos[2\psi(\varphi)] }{ 1 - \alpha\cos[2 \psi(\varphi)] } },
\end{equation}
where $\psi(\varphi) \equiv \theta_{\rm ssd}(\varphi)-\varphi$ and $p$ is a constant fixed by the circuitation condition $\int_0^{2\pi}\rmd\varphi\, \partial_\varphi\theta_{\rm ssd}=2\pi s$. Equation~\eref{implnem} can in principle be solved iteratively at any order in $\alpha$. 
At ${\cal O}(1)$, the solution is simply that of the isotropic theory, namely, $\theta_{\rm ssd}(\varphi) = s\varphi + \theta_{\rm c} + {\cal O}(\alpha)$, where $\theta_{\rm c}$ is an integration constant.
At linear order in $\alpha$, the solution becomes
\begin{equation}
   \label{approxstaticnem}
    \partial_\varphi\theta_{\rm ssd}(\varphi)= s + \frac{\alpha}{2}\frac{s(s-2)}{s-1}\cos\left(2(s-1)\varphi + 2\theta_{\rm c}\right)+{\cal O}(\alpha^2).
\end{equation}
Inserting this expression in Eq.~\eref{eq_barmu_der_2} with $s = \pm \frac{1}{2}$ and integrating over $\varphi$,
we find that $\nu_{ij}=\bar{\nu}_{ij}=\frac{\pi}{4} \delta_{ij}$.
Therefore, at linear order in $\alpha$ oppositely charged defects have equal isotropic mobilities, while corrections are expected at higher order in perturbation~\cite{Sven2003PRL,Brugues2008PRL} (see also Sec.~\ref{sec_mob_alpha2}).

\begin{figure}[t!] 
    \centering\includegraphics[width=.8\linewidth]{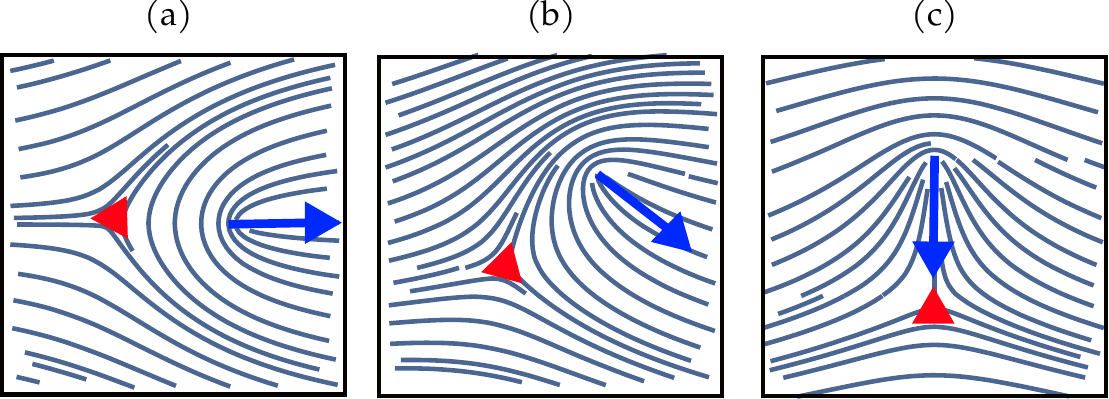}
    \caption{Configurations of a pair of nematic defects corresponding to $\theta_q=0$ (a), $\theta_q=\frac{\pi}{4}$ (b), and $\theta_q=\frac{\pi}{2}$ (c).
    The grey lines indicate the local orientation of the nematic field. The polarization $\hat{\bi{p}}$ of the $+\frac{1}{2}$ defect is shown with the blue arrows, while red triangles are used to indicate the position of the $-\frac{1}{2}$ defect.  
    }
    \label{fig_sketch_ani_defects}
\end{figure}

The evaluation of the force term in~\eref{eq_logaexp} is less straightforward and calculation details are presented in~\ref{anisforce}. In summary, we express the static orientation field created by the two defects as $\theta_{\rm sb}(\x) = \theta_0(\x) + \alpha \theta_1(\x) + {\cal O}(\alpha^2)$, where $\theta_0(\x)$ solves the linear problem ($\alpha = 0$) 
and $\theta_1(\x)$ is calculated perturbatively.
However, the formal expression of $\theta_1$ still involves intricate integrals. We thus use the fact that the integration contour on the r.h.s.\ of Eq.~\eref{eq_logaexp} is arbitrary, and
integrate over the bisector of the segment formed by the two defects. This choice conveniently allows to express all integrals independently of any model parameters, such that they reduce to numerical constants. After numerically computing these coefficients, we finally obtain for the vector $\q = \frac{1}{2}(\q_+ - \q_-)$
\begin{equation}
    \label{anismotion}
    \ln\left(\frac{\lambda}{a}\right) \dot{\q} = -\frac{1}{q}\left[\hat{\q} - \frac{4 \alpha}{3}\sin(2\theta_q)\eps\hat{\q} \right],
\end{equation}
where $\theta_q$ denotes the angle between $\q$ and the direction of the nematic order parameter at infinity parametrized by $\theta_{\rm c}$.
Comparing Eqs.~\eref{anismotion} and~\eref{eq_veff_2def}, we see that elastic anisotropy generates, at the linear order in $\alpha$, a tangential force that essentially aligns $\q$ along the background nematic order orientation for $\alpha < 0$, or orthogonal to it for $\alpha > 0$. To give a physical interpretation of this force, we show in Fig.~\ref{fig_sketch_ani_defects} that varying $\theta_q$ amounts to changing the relative orientation of the two defects.
The tangential force in Eq.~\eref{anismotion} then reflects the energetic cost of misaligned defects due to elastic anisotropy, and favors configurations with $\theta_q = \frac{\pi}{2}$ or $0$.

\begin{figure}[t!] 
    \centering
    \includegraphics[width=.8\linewidth]{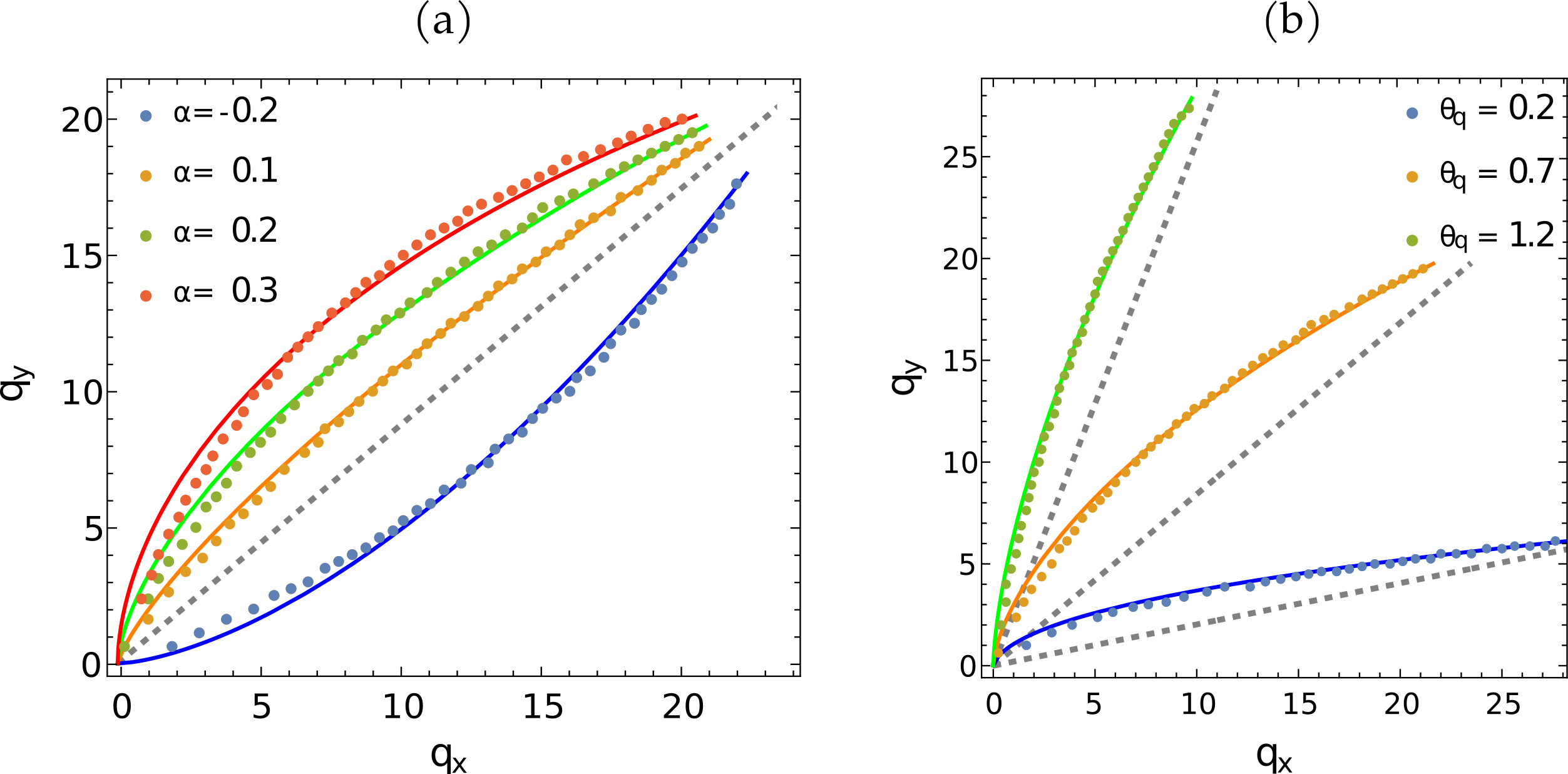}
    \caption{
    Trajectories of the vector $\q$ obtained from numerical integration of Eq.~\eref{anismotion} (full lines) and direct simulations of the annihilation of isolated $\pm\frac{1}{2}$ defect pairs in the full theory~\eref{fullnematic} (symbols). 
    Panel (a) shows trajectories at fixed $\theta_q = 0.7$ for various values of $\alpha$, while panel (b) corresponds to $\alpha = 0.2$ and different values of $\theta_q$ (rad).
    In (a,b) the dashed lines correspond to the expected trajectories at $\alpha = 0$.
    }
\label{bigcurves}
\end{figure}

To confirm the predictions of Eq.~\eref{anismotion}, we performed numerical simulations of the dynamics described by
\begin{eqnarray}
\label{fullnematic}
\partial_t \Q_{ij} & = -\frac{\delta \calF_{\rm an} }{\delta \Q_{ij}} + \frac{\delta_{ij}}{2} {\rm Tr}\left[\frac{\delta \calF_{\rm an} }{\delta \bQ} \right], \\
\calF_{\rm an} & = \int \rmd^2\x \, \left[ \frac{1}{2}\left(\delta_{kl} + \sqrt{2}\alpha \Q_{kl}\right)(\partial_k \Q_{ij})(\partial_l \Q_{ij}) + \chi^2\left( 1- {\rm Tr}[\bQ^2] \right)^2 \right], \nonumber
\end{eqnarray}
for the full nematic order parameter $\bQ$, where $\chi \sim a^{-1}$ sets the size of the defect core. $\calF_{\rm an}$ is the simplest free energy that admits~\eref{simplefrank} as a bulk theory, which can be verified by setting $\rho = 1$ in Eq.~\eref{fullnematic}.
Figure.~\ref{bigcurves} shows trajectories of the vector $\q$ obtained from initially prepared $\pm\frac{1}{2}$ defect pairs at various values of $\alpha$ and initial orientation $\theta_q$ (details on numerical simulations are given in~\ref{app_numerics}).
Note that the mobility in Eq.~\eref{anismotion} only affects the speed of the defects, and not their trajectories.
Hence, studying the trajectories we can compare the results obtained from the integration of Eq.~\eref{anismotion} and simulations of the full model~\eref{fullnematic} without any fitting parameter. 
In qualitative agreement with the predictions of Eq.~\eref{anismotion}, we find that for $\alpha \ne 0$ they annihilate following curved trajectories, as a result of the tangential component of the force between misaligned defects. We observe that the magnitude of the curvature increases with $\alpha$.
Quantitatively, the curves shown in Fig.~\ref{bigcurves} reveal a good agreement between theory and simulations, 
whereas deviations appear at larger $\alpha$ and close to the annihilation point, which we interpret as the breakdown of our perturbative approach and the slow defect approximation, respectively. 

\subsubsection{Anisotropic mobility of $+\frac{1}{2}$ defects}
\label{sec_mob_alpha2}

We showed previously that at the linear order in $\alpha$ the defect mobility is not affected by elastic anisotropy.
Data obtained at stronger anisotropy from numerical simulations~\cite{Sven2003PRL} or in experiments~\cite{Brugues2008PRL}, however, suggest that higher order corrections will affect the mobilities of the defects.
In this section, we therefore calculate the leading order contribution of elastic anisotropy to the defect mobility, which arise at ${\cal O}(\alpha^2)$.
At this order, the orientation field obtained from \eref{implnem} is given by
\begin{eqnarray}
\fl \qquad\quad 
\partial_\varphi\theta_{\rm ssd}(\varphi) = & s + \frac{\alpha}{2}\frac{s(s-2)}{s-1}\cos\left(2(s-1)\varphi + 2\theta_{\rm c}\right) \nonumber \\
\label{eq_sec_ord_theta}
& + \frac{\alpha^2(5 s^3-20 s^2+24 s-8
)}{16(s-1)^3}\cos(4(s-1)\varphi + 4\theta_{\rm c}) + {\cal O}(\alpha^3).
\end{eqnarray}

We use Eq.~\eref{eq_barmu_der_2} to calculate the defect friction. 
Splitting it into isotropic and anisotropic parts, we obtain
\begin{equation} \label{mobsplit}
{\nu}_{ij} = \frac{1}{2}\int_0^{2\pi} \rmd\varphi \, (\partial_{\varphi}\theta_{\rm ssd}(\varphi))^2 \left[ \delta_{ij} + \sqrt{2}\Q_{ij}(\varphi) \right]. 
\end{equation}
For the isotropic part, we obtain after some algebra
\begin{equation*}
\label{isopart}
\frac{1}{2}\int_0^{2\pi} \rmd\varphi \, (\partial_{\varphi}\theta_{\rm ssd}(\varphi))^2=s^2\pi\left(1+\alpha^2\frac{(s-2)^2}{8(s-1)^2}\right) + {\cal O}(\alpha^3).
\end{equation*}
In addition, the anisotropic part of the friction is found to be
\begin{equation*}
\label{isopart}
\frac{1}{\sqrt{2}}\int_0^{2\pi} \rmd\varphi \, (\partial_{\varphi}\theta_{\rm ssd}(\varphi))^2 \Q_{ij}(\varphi) = \frac{3\pi \alpha^2}{8\sqrt{2}}\Q_{ij}(2\theta_{\rm c})\delta_{s,\frac{1}{2}} + {\cal O}(\alpha^3).
\end{equation*}
Hence, only $s = +\frac{1}{2}$ defects have an anisotropic \Correction{friction tensor}.
As we argue in~\ref{mobilandsymm}, we expect this result to hold at all orders in $\alpha$.

We now note that $\theta_{\rm c}$ is defined in the previous section as the background orientation of the nematic order.
Conversely, by rotational invariance of the problem $\theta_{\rm c}$ may define the direction of the comet-shaped $+\frac{1}{2}$ defect.
Namely, it is straightforward to show from~\eref{eq_sec_ord_theta} and for $s =\frac{1}{2}$ that rotating $\theta_{\rm c}$ by an angle $\vartheta$ 
amounts to rotate the defect solution by an angle $2\vartheta$.
Hence, we define the $+\frac{1}{2}$ defect polarization as $\hat{\bi{p}} \equiv (\cos2\theta_{\rm c} , \sin2\theta_{\rm c})$ (see the blue arrows in Fig.~\ref{fig_sketch_ani_defects}), such that we finally obtain
for the two types of nematic defects
\begin{eqnarray} 
\label{eq_nu_plus}
\bnu_{\frac{1}{2}} = \frac{\pi}{4}\left(1+\frac{21}{8}\alpha^2\right) \hat{\bi{p}}\hat{\bi{p}} + \frac{\pi}{4}\left(1 - \frac{3}{8}\alpha^2\right)( \bi{I} - \hat{\bi{p}}\hat{\bi{p}}), \\
\label{eq_nu_minus}
\bnu_{-\frac{1}{2}} = \frac{\pi}{4}\left(1+\frac{25}{72}\alpha^2\right) \bi{I}.
\end{eqnarray}
Comparing the coefficients of Eq.~\eref{eq_nu_plus}, we thus find that the $+\frac{1}{2}$ defect experiences a larger friction 
in the direction longitudinal to its comet-like shape, whereas elastic anisotropy reduces the strength of the friction in the transverse direction.
Conversely, elastic anisotropy renormalizes the friction of the $-\frac{1}{2}$-charged defect upwards in all directions.

\section{Domain walls and disclination lines} \label{sec_lines}

We now illustrate how the variational approach outlined in the previous section can be extended to describe defects with different geometries. 
We start by discussing the simple case of domain walls in the presence of $\mathbb{Z}_2$ broken symmetry, and then extend our derivation to disclination lines.

\subsection{Domain walls: the Allen-Cahn equation} \label{sec_dw}

\begin{figure}[t!] 
\centering
    \centering\includegraphics[width=.37\linewidth]{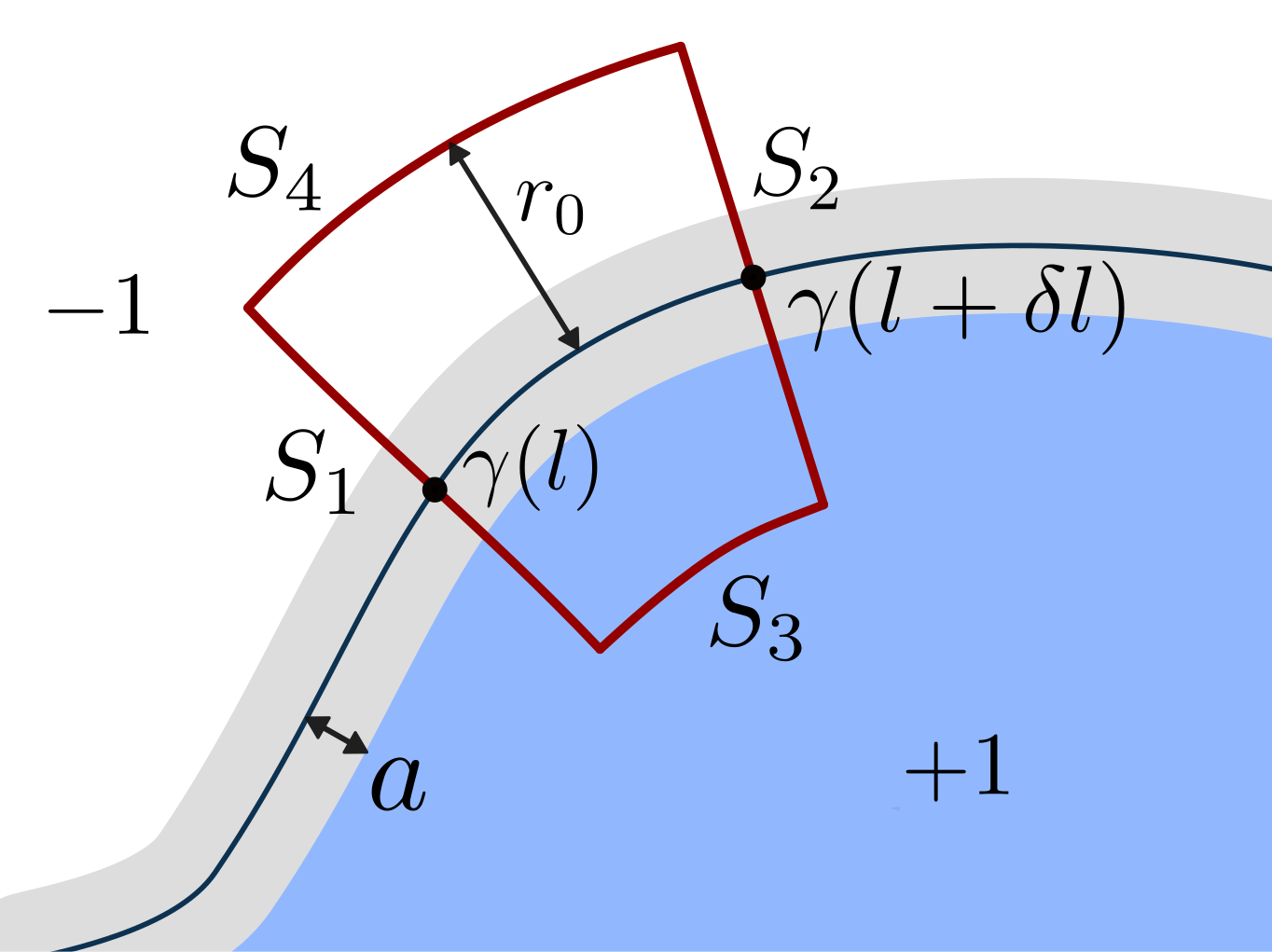}
    \caption{A domain wall (grey-shaded region) separating two homogeneous regions with $\bar\phi = \pm 1$.
    The choice of boundary to calculate the variation is shown in red. The surfaces $S_{1,2}$ define the infinitesimal segment, while $S_{3,4}$ show the matching region.}
    \label{fig:domainvariation}
\end{figure}

Domain walls form in systems described by a scalar order parameter $\phi(\x ,t)$ that accounts for spontaneously broken $\mathbb{Z}_2$ symmetry. Considering units for which $\phi$ equilibrates to the values $\bar{\phi} = \pm 1$, they correspond to thin interfaces separating homogeneous regions where $\phi$ takes opposite signs (Fig.~\ref{fig:domainvariation}).
A minimal continuous description for this class of system is given by the following free energy
\begin{equation}
    \label{GLandau}
    \Fr = \int \rmd^2\x \, \left[ \frac{1}{2}|\nabla \phi|^2 + V(a,\phi) \right],
\end{equation}
where $V(a,\phi)$ is an arbitrary bulk potential that depends on a microscopic scale $a$ setting the thickness of the domain walls. In particular, a common example is the Landau potential $V(a,\phi) = a^{-2}(1-\phi^2)^2$. The derivation below relies on the same assumptions~(\ref{ass:1}-\ref{ass:4}) that were extensively used in Sec.~\ref{sec_exact_var}. Importantly, we will work in the limit $a \to 0$ ensuring a separation of scales between the domain wall thickness and any macroscopic length.

We consider a domain wall described by the curve $\vg(l,t)$ on the plane, where $l$ sets the curve parametrization and $t$ is the time.
In what follows we will denote as $\vg'$ and $\dot{\vg}$ derivatives of $\vg$ with respect to $l$ and $t$, respectively. 
The derivation of the equation of motion for an infinitesimal segment $[\vg(l,t);\vg(l+\rmd l,t)]$ roughly follows that outlined in Sec.~\ref{sec_exact_var}.
However, as the object we now describe is one-dimensional, we need to choose an appropriate volume $\calV$ to delimit the matching region.
Given a point $\vg(l,t)$, we define $\calV$ as a tetragon enclosing the domain wall (as sketched in Fig.~\ref{fig:domainvariation}).
The edges $S_1$ and $S_2$ cross the interface orthogonally respectively at $\vg(l,t)$ and $\vg(l+\rmd l,t)$, and extend on both sides up to a distance $r_0 \gg a$. 
In addition, the other two edges $S_3$ and $S_4$ close the curve and are both fully in the matching region.

Following similar steps as in Sec.~\ref{sec_exact_var},
and noting that 
inside of $\calV$, $\partial_t\phi=-\dot{\vg}(l,t)\cdot\nabla\phi$, we obtain
\begin{equation} \label{eq_gamma_walls0}
    \dot{\gamma}_j(l,t)\int_{\calV}\rmd^2\x\,(\partial_i\phi)(\partial_j\phi) = - \sum_{k=1}^4 \int_{S_k}\rmd S_j T_{ji} ,
\end{equation}
where $\bi{T}$ is the stress tensor associated with $\calF$ and the $\rmd\bi{S}$ vectors point to the outside of $\calV$.
As the integrals over $S_{3,4}$ are evaluated in the bulk of the system where the field is homogeneous and takes values $\pm 1$ (Fig.~\ref{fig:domainvariation}), \Correction{the stress tensor identically vanishes on both of these integration regions, which do not contribute to the force.}
To calculate the contributions from the integrals over $S_{1,2}$, we assume that the curve $\vg$ is sufficiently smooth such that its local curvature satisfies $\kappa(l,t) a \ll 1$.
Denoting $\hat{\bi{t}}(l,t)$ and $\hat{\bi{n}}(l,t)$ the local tangential and normal vectors to $\vg$, 
it is clear that 
\begin{equation*}
    |\hat{\bi{t}}\cdot\nabla\phi| \propto \kappa \ll
    |\hat{\bi{n}}\cdot\nabla\phi| \propto a^{-1}.
\end{equation*}
In what follows, we thus only retain the dominant contribution $\partial_{\hat{\bi{n}}}\phi \equiv \hat{\bi{n}}\cdot\nabla\phi$ to $\nabla\phi$,
and neglect $\partial_{\hat{\bi{t}}}\phi \equiv \hat{\bi{t}}\cdot\nabla\phi$.
Moreover, as the edges $S_{1,2}$ are orthogonal to $\vg$ in $l$,
we have $\rmd\bi{S}_{1} = - \rmd x_{\rm n} \, \hat{\bi{t}}(sl,t)$ and $\rmd\bi{S}_{2} = \rmd x_{\rm n} \, \hat{\bi{t}}(l+\rmd l,t)$, where hereafter $x_{\rm n}$ denotes the running coordinate normal to the interface.
Then,
\begin{equation}
\label{firstsurf}
\rmd S_{1,j}T_{ji} = - \rmd x_{\rm n} \, \hat{t}_j 
\left(\partial_i \phi \,\frac{\partial F}{\partial (\partial_j \phi)}-\delta_{ij}F \right) = \rmd x_{\rm n} \, \hat{t}_i F,    
\end{equation}
where $F$ is the free energy density associated with~\eref{GLandau},
while the first contribution on the r.h.s.\ was eliminated noting that the stress tensor of a scalar theory is always symmetric.
After expressing $\rmd{S}_{2,j} {T}_{ji}$ in a similar way, we obtain
\begin{equation}
\fl \qquad -\left[ \int_{S_1} + \int_{S_2} \right] \rmd S_j T_{ji} = 
    \left[\hat{t}_i(l + \rmd l,t) - \hat{t}_i(l ,t)\right]
    \int_{-\infty}^{+\infty}\rmd x_{\rm n}\, F 
    \equiv  \sigma \rmd l \, \hat{t}_i'(l ,t),
\end{equation}
where the $\pm \infty$ integration boundaries follow from the fact that $r_0 \gg a$ and $F = 0$ in the bulk phases.
The constant $\sigma$ represents the free energy per unit length of the interface, and thus corresponds to the surface tension.

We now calculate the \Correction{friction tensor} in~\eref{eq_gamma_walls0}. At the leading order in $\rmd l$,
\begin{equation}
\label{inmob}
\int_{\calV}\rmd^2\x\,(\partial_i\phi)(\partial_j\phi)
= \rmd l\, |\vg'(l,t)| \hat{n}_i\hat{n}_j \int^{\infty}_{-\infty}\rmd x_{\rm n}\, (\partial_{\hat{\bi{n}}}\phi)^2 ,
\end{equation}
where we have again used the fact that $\partial_{\hat{\bi{t}}}\phi$ is subdominant and that $\nabla\phi$ vanishes in the bulk.
Defining the friction as $\zeta = \int^{\infty}_{-\infty}\rmd x_{\rm n}\, (\partial_{\hat{\bi{n}}}\phi)^2$, we finally recover the Allen-Cahn equation~\cite{Bray2002review}
\begin{equation}
    \label{walldyn}
    \zeta \, \n \cdot \dot{\bgamma}(l,t)= \sigma\frac{\kappa(l,t)}{| \vg'|},
\end{equation}
where $\kappa = |\hat{\bi{t}}'|$ is the local curvature of the interface.
As expected, we thus find that the transverse velocity of the interface is set by the local curvature.
The surface tension and friction are moreover fully determined by the interface profile, which can be obtained from~\eref{GLandau} by solving
$\partial_{\n\n}^2\phi - \partial_\phi V(a,\phi) = 0$.
Multiplying both sides by $\partial_{\n}\phi$ and integrating across the interface, it is straightforward to show that $\frac{1}{2}(\partial_{\n}\phi)^2 - V(a,\phi) = 0$.
It then follows that
\begin{equation}
    \sigma = \int_{-\infty}^{\infty}\rmd x_{\rm n}\, \left[ \frac{1}{2}(\partial_{\n}\phi)^2 + V(a,\phi) \right] = 
    \int_{-\infty}^{\infty}\rmd x_{\rm n}\, (\partial_{\n}\phi)^2  = 
    \zeta,
\end{equation}
which is the well known relationship between surface tension and friction for domain walls~\cite{Bray2002review}.
Defining $\dot{\gamma}_\perp(l,t) \equiv \n\cdot\dot{\vg}$, and choosing the arc-length parametrization for $\vg$, we end up with the compact formula
\begin{equation} \label{eq_AC2D}
    \dot{\gamma}_\perp(l,t) = \kappa(l,t).
\end{equation}
The generalization of Eq.~\eref{eq_AC2D} to three dimensions, where domain walls take the from of two-dimensional surfaces, 
is straightforward.
Parametrizing the domain wall with the coordinates $\bi{l}$, any point $\bLambda(\bi{l},t)$ on the manifold then evolves as
\begin{equation} \label{eq_domainwall_3D}
    \dot{\Lambda}_\perp(\bi{l},t) \equiv \hat{\bi{n}}(\bi{l},t)\cdot\dot{\bLambda}(\bi{l},t) = -2 H(\bi{l},t),
\end{equation}
where $\hat{\bi{n}}$ is the vector normal to the interface (assuming that the surface is closed, $\hat{\bi{n}}$ points to the outside) and $H$ is its mean curvature.
Equation~\eref{eq_domainwall_3D} is known as the Allen-Cahn equation~\cite{ALLEN1979}, and states that the domain wall velocity is only determined by its local mean curvature since the friction and the surface tension are proportional to each other.

\subsection{Disclination lines in three dimensions} \label{sec_discl_lines}

\subsubsection{The general equation of motion for disclination lines}

We now discuss the situation where the order parameter $\bphi$ is two-dimensional and evolves in a three-dimensional space.
In this scenario, defects take the form of one-dimensional manifolds around which the circulation of the order parameter orientation is topologically constrained, and that are usually referred to as vortex- or disclination lines~\cite{Aranson2022RMP}.
Similar structures are for example found in superfluid Helium~\cite{Fetter2001BEC} or in the displacement field of sheared glasses~\cite{Wu2023NatCom}.
Similarly to point defects, the motion of disclination lines has been mostly studied in the context of the classical Ginzburg-Landau theory via matched asymptotics~\cite{Rubinstein1991} or using a variational approach based on the Rayleigh dissipation function~\cite{Kawasaki1983linedef}.
Here, we first show how to obtain the equation of motion for disclination lines arising in a system whose dynamics minimizes an arbitrary free energy, while the application to the Ginzburg-Landau framework is presented in a second part.

\begin{figure}[!t] 
\centering
    \centering\includegraphics[width=.3\linewidth]{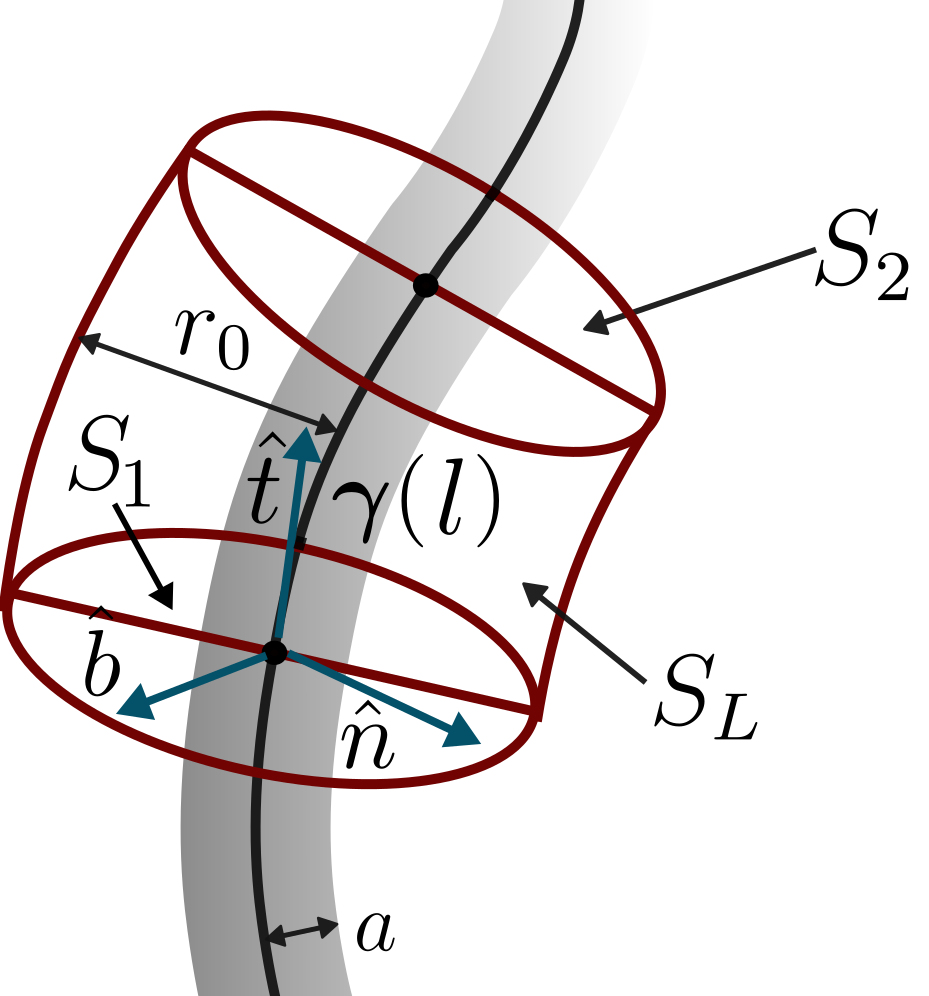}
    \caption{Choice of boundary for defect lines.}
    \label{defbound}    
\end{figure}

As before, we denote by $\vg(l,t)$ the vector defining the defect line, and choose an adequate parametrization such that $|\vg'(l,t)| = 1$.
In three dimensions, the curve $\vg(l,t)$ is locally defined by the Frenet-Serret frame
\begin{equation}
    \hat{\bi{t}} = \vg', \qquad
    \hat{\bi{n}} = \kappa^{-1} \hat{\bi{t}}', \qquad
    \hat{\bi{b}} = \hat{\bi{t}} \times \hat{\bi{n}},
\end{equation}
such that $\hat{\bi{t}}$ is tangent to $\vg$, $\hat{\bi{n}}$ and $\hat{\bi{b}}$ are respectively the normal and binormal unit vectors, and $\kappa$ denotes the local curvature of the curve.
As before, we assume a separation of scales between the core and bulk physics, such that we work in the limit where the core radius $a$ satisfies $\kappa a \ll 1$.
Given the geometry of the problem, the natural choice for the volume $\calV$ is an infinitesimal cylinder with top and bottom surfaces $S_1$ and $S_2$ orthogonal to \Correction{the curve} $\vg$ respectively in $l$ and $l + \rmd l$, while the whole lateral surface $S_{\rm L}$ lies in the matching region (see a sketch in Fig.~\ref{defbound}).
The application of the free energy variation method in this case combines the results obtained in the previous sections for point-defects and domain walls. 
We first evaluate the defect friction under the assumption~(\ref{ass:3}) of a rigid core. When the local radius of curvature of $\vg$ is large as compared to the size of the core, we obtain 
\begin{eqnarray}
\label{eq:frictionequalities}
    \zeta_{ij} & = \int_{\calV}\rmd^3\x \, (\partial_i\bphi) \cdot (\partial_j\bphi) \nonumber \\
     & \simeq \frac{1}{2} \rmd l (\delta_{ij} - \hat{t}_i \hat{t}_j)\int_{{\cal D}_{r_0}} \rmd^2\x_{\rm n}\,  (\partial_k\bphi) \cdot (\partial_k\bphi) \nonumber\\
    & \equiv \rmd l (\delta_{ij} - \hat{t}_i \hat{t}_j)\zeta\left(\frac{r_0}{a}\right), \label{eq_mob_line_defect}
\end{eqnarray}
at leading order in $\rmd l$, where ${\cal D}_{r_0}$ denotes the disk of radius $r_0$ orthogonal to $\hat{\bi{t}}$.
The second equality is obtained by 
neglecting the part of $\nabla\bphi$ tangential to $\hat{\bi{t}}$ and assuming that the \Correction{tensor coming from the integral on the r.h.s.\ is isotropic on the plane normal to $\vg(l)$}. \Correction{This assumption is motivated by observing that, in the limit of large loop curvature, the integral in the second line of~\eref{eq:frictionequalities} has properties analogous to the friction integral studied for topological defects in Sec.~\ref{sec_top_def}, which we have shown to lead to isotropic friction tensor in most situations.} 
Analogously to domain walls, the integration of the stress tensor over the surfaces $S_1$ and $S_2$, \Correction{respectively orthogonal to $\vg(l)$ and $\vg(l+dl)$}, gives the surface tension contribution:
\begin{equation} \label{eq_sigma_defect}
    -\left[\int_{S_1} + \int_{S_2}\right] d S_j T_{ji} = \sigma\left(\frac{r_0}{a}\right) \hat{t}_i'\rmd l = \sigma\left(\frac{r_0}{a}\right)\kappa \hat{n}_i\rmd l,
\end{equation}
where $\sigma = \int_{{\cal D}_{r_0}}\rmd^2\x_{\rm n} \, F$. 
At this stage, we can already note that, contrary to the case of domain walls, both the friction and surface tension depend on the matching variable $r_0$ and the core size $a$. 
In fact, we will show later that they are both logarithmically divergent in $r_0$, similarly to the situation studied in Sec.~\ref{sec_def_GL} for point-defects.

We now calculate the contribution of the lateral surface $S_{\rm L}$. \Correction{Contrary to domain walls, the bulk stress tensor does not vanish here because of the presence of the long-range modulations of $\theta$ which mediate interactions between the defect lines.} 
Denoting $(\hat{\vr},\hat{\bvarphi},\hat{\bi{t}})$ as the local cylindrical frame centred in $\bgamma$, 
the surface element on $S_{\rm L}$ can be expressed as $\rmd{S}_i = - r_0 \rmd\varphi \rmd l \epsilon_{ijk}\hat{t}_j \hat{\varphi}_k$ where $\epsilon_{ijk}$ is the 3D totally antisymmetric Levi-Civita tensor.
Hence,
\begin{equation} \label{eq_SL}
    - \int_{S_{\rm L}} \rmd {S}_j \, {T}_{{\rm bulk},ji} 
    = r_0\rmd l \epsilon_{klj} \hat{t}_k \int_{0}^{2\pi} \rmd \varphi \, \hat{\varphi}_l {T}_{{\rm bulk},ji},
\end{equation}
where we have used the fact that the integration is done in the matching region to substitute $\bi{T}$ by its bulk counterpart.
Combining Eqs.~(\ref{eq_mob_line_defect},\ref{eq_sigma_defect},\ref{eq_SL}), and denoting $\dot{\vg}_\perp \equiv (\bi{I} - \hat{\bi{t}}\hat{\bi{t}})\dot{\vg}$,
we finally obtain
\begin{equation}
\label{lineq}
    \zeta\left(\frac{r_0}{a}\right)\dot{\gamma}_{\perp,i}
    =  \sigma\left(\frac{r_0}{a}\right)\kappa\hat{n}_i
    - r_0 \epsilon_{klj} \hat{t}_k \int_{0}^{2\pi} \rmd \varphi \, \hat{\varphi}_l {T}_{{\rm bulk},ji}.
\end{equation}
Similarly to point-defects, Eq.~\eref{lineq} explicitly depends on the matching variable $r_0$. 
As $r_0$ is arbitrary, it must simplify when a bulk free energy is specified.

\subsubsection{Isotropic systems: the large $\ln(a)$ expansion}
We now consider a bulk free energy analogous to that used in Sec.~\ref{sec_def_GL}:
\begin{equation}
    \calFbulk=\frac{1}{2}\int \rmd^3\x \, |\nabla\theta|^2.
\end{equation}
The corresponding dynamical evolution of the orientation field $\theta(\x,t)$ is given by the three-dimensional diffusion equation. 
As this equation is linear, it can formally be solved using the approach presented in~\cite{romano2023},
albeit leading to a more complicated solution than in two-dimensions. 
Here, we instead note that taking the derivative of the friction coefficient and the surface tension with respect to $r_0$, we obtain
\begin{eqnarray}
\label{mobandfreeenergy}
    \frac{\rmd \zeta}{\rmd r_0} = \frac{r_0}{2} \int_0^{2\pi}\rmd\varphi\, |\nabla\theta|^2,\qquad
    \frac{\rmd \sigma}{\rmd r_0} = r_0 \int_0^{2\pi}\rmd\varphi \, F_{\rm bulk} 
    = \frac{\rmd \zeta}{\rmd r_0},
\end{eqnarray}
where the integrals are evaluated at the leading order in $a$ from the static straight line defect solution.
This solution simply corresponds to an extension of the point-defect solution into the third dimension:
\begin{equation}
\label{staticlowcurv}
    \nabla\theta_{\rm ssd}(\x)= s\,\frac{\hat{\bvarphi}}{r},
\end{equation}
where $r$ is the minimal distance between $\x$ and $\vg$.
Plugging~\eref{staticlowcurv} in Eq.~\eref{mobandfreeenergy} then straightforwardly leads to $\zeta \simeq \sigma \simeq \pi s^2 \ln(r_0/a)$.
We now take $a \to 0$ and perform a low mobility expansion as was done in Sec.~\ref{sec_loga_exp}.
Therefore, we calculate the force term in~\eref{lineq} via the static line defect solution which for general curves is given by the Biot-Savart law:
\begin{equation} \label{eq_BS}
    \nabla\theta_{\rm sd}(\x) = \frac{s}{2} \int \rmd \vg \times \frac{\x - \vg}{|\x - \vg|^3}.
\end{equation}
In particular, in the specific case of a straight disclination line aligned along $\hat{\bi{z}}$, the expression~\eref{eq_BS} simplifies into~\eref{staticlowcurv}.
Since we integrate the stress tensor over a circle of radius $r_0 \to 0$ enclosing the line around $\vg$, 
we write the phase field for $\x \to \vg$ as $\nabla\theta = \nabla\theta_{\rm d} + \nabla\theta_{\rm c}$.
The discontinuous part is obtained by expanding~\eref{eq_BS} in the vicinity of $\vg$, giving $\nabla\theta_{\rm d}(\x) = \nabla\theta_{\rm ssd}(\x) - \kappa s \ln(\kappa r_0) \hat{\bi{b}}/2$, while the continuous part $\theta_{\rm c}$ accounts for other singularities or externally imposed boundary conditions.
After evaluating the integral of the stress tensor explicitly, we obtain
\begin{equation}
    \label{slowandstraexp}
    \ln\left(\frac{L}{a}\right)\dot{\vg}_\perp
    =  \kappa\ln\left(\frac{1}{\kappa a}\right)\hat{\bi{n}}
    - \frac{2}{s}\, \hat{\bi{t}} \times \nabla\theta_{\rm c}(\vg) .
\end{equation}
While the matching scale $r_0$ in the expression of the surface tension was simplified due to the curvature contribution to $\nabla\theta_{\rm d}$, 
as previously discussed for slow defects the $r_0$ dependency of the friction can be replaced by any arbitrary macroscopic scale $L$.
Equation~\eref{slowandstraexp} can be interpreted as the generalization of~\eref{isomotion} (in the large $\ln(a)$ limit) to the case where the singularities extend along a third dimension. 
Importantly, it includes an additional term on the r.h.s.\ leading to the collapse of disclination lines with finite curvature, while the Peach-Koehler-type force describes how distinct lines interact.

We note that since the friction coefficient and the surface tension share the same scaling with $\ln(a)$, the importance of the interaction term to the dynamics of disclination lines depends on their curvature.
Indeed, assuming $\kappa$ to be ${\cal O}(1)$ implies that the line velocity is also ${\cal O}(1)$, while the interaction term on the r.h.s.\ of~\eref{slowandstraexp} is ${\cal O}(\ln^{-1}(a))$. 
Hence, the dynamics is dominated by the tension and, after taking $L \approx \kappa^{-1}$, we recover an equation similar to that ruling the motion of domain walls:
\begin{equation}
\label{lineq_highk}
    \dot{\vg}_\perp
    = \kappa\hat{\bi{n}}, \qquad (\kappa = {\cal O}(1)).
\end{equation}
Expressions similar to~\eref{lineq_highk} have been derived by several authors who focused on the dynamics of isolated line defects~\cite{Rubinstein1991,Aranson2022RMP}.

On the other hand, in the presence of several line defects with low curvature (namely, if $\kappa = {\cal O}(\ln^{-1}(a))$, the interaction term in~\eref{slowandstraexp} becomes relevant.
In the quasi-static approximation and assuming open boundaries, the term $\nabla\theta_{\rm c}(\vg_\alpha)$ appearing in the equation of motion of the defect $\alpha$ can thus be generally expressed as $\nabla\theta_{\rm c}(\vg_\alpha) = \sum_{\beta \ne \alpha} \nabla\theta_{{\rm sd},\beta}(\vg_\alpha)$, where $\theta_{{\rm sd},\beta}$ corresponds to the orientation profile generated by the defect $\beta$.

In the specific case where all disclination lines are straight and aligned along $\hat{\bi{z}}$, it is straightforward to check from~\eref{eq_BS} that~\eref{slowandstraexp} reduces to the description of the two-dimensional Coulomb gas~\cite{Pismen1991PhysicaD}.
Conversely, for straight disclination lines with arbitrary orientations, Eq.~\eref{slowandstraexp} takes the form
\begin{equation} \label{eq_straight_deflines}
    \ln\left( \frac{L}{a} \right)\dot{\vg}_{\perp,\alpha} = \sum_{\beta \ne \alpha}\frac{s_\beta}{s_\alpha }\;\frac{
    \hat{\vr}_{\alpha \beta} (\hat{\bi{t}}_\alpha \cdot \hat{\bi{t}}_\beta)
    - \hat{\bi{t}}_\beta (\hat{\bi{t}}_\alpha \cdot \hat{\vr}_{\alpha \beta})}{r_{\alpha \beta}},
\end{equation}
where $\vr_{\alpha \beta} \equiv \frac{1}{2}(\vg_\alpha - \tilde{\vg}_{\alpha \beta})$ and $\tilde{\vg}_{\alpha \beta} \equiv {\rm argmin}_{\vg_\beta}(|\vg_\alpha - \vg_\beta|)$.
The second contribution to the r.h.s.\ of~\eref{eq_straight_deflines} aligns or anti-aligns the direction of the lines $\alpha$ and $\beta$ (parametrized by their tangent vector) whenever $s_\alpha s_\beta < 0$ or $s_\alpha s_\beta > 0$, respectively.
Hence, due to the first term the effective interaction between disclination lines is always attractive regardless of their respective charges.

\subsubsection{Disclination loops} \label{sec_disc_loops}
Another type of structure of interest are loop singularities.
To study their dynamics, we consider a circular ring $\alpha$ centred at position $\bi{X}_\alpha$ with radius $R_\alpha$. We define the associated dipole moment as
\begin{equation}
   \label{eq:magnmom}
   \bi{m}_\alpha \equiv \frac{s_\alpha}{4}\oint \, \vg_\alpha \times \rmd\vg_\alpha = \frac{\pi s_\alpha R_\alpha^2}{2}\, \hat{\bi{b}}_\alpha,
\end{equation}
where $\hat{\bi{b}}_\alpha$ is the binormal unit vector orthogonal to the \Correction{loop} plane. 
Taking the time derivative of Eq.~\eref{eq:magnmom} and using~\eref{slowandstraexp}, we find that the loop radius and orientation obey
\begin{eqnarray}
\label{eq:alint1}
    \dot{R}_\alpha & =-\frac{1}{R_\alpha}-\frac{2}{s_\alpha \ln\left( R_\alpha/a \right)} \left(\hat{\bi{b}}_\alpha\cdot \nabla\theta_{\rm c}\right),\\
\label{eq:alint2}
    \dot{\hat{\bi{b}}}_\alpha & =\frac{2}{s_\alpha R_\alpha \ln\left( R_\alpha/a\right)}
    \hat{\bi{b}}_\alpha\times\left(\hat{\bi{b}}_\alpha\times\nabla\theta_{\rm c}\right),
\end{eqnarray}
while the dynamics of $\bi{X}_\alpha$ can be obtained by simply averaging~\eref{slowandstraexp} over the ring contour.
Note that we have chosen to replace $L$ with $R_\alpha$ in~(\ref{eq:alint1},\ref{eq:alint2}), as here $1/R_\alpha$ sets the natural scale for the line velocity.
We first consider the case where the loop is subject to a uniform orientation gradient $\nabla\theta_{\rm c} = \bi{k}$, as was done in Sec.~\ref{sec_def_const_vel}.
It is then straightforward to check that $\dot{\bi{X}}_\alpha = \bf{0}$.
Moreover, we conclude from Eq.~\eref{eq:alint2} that the ring anti-aligns (aligns) with $\bi{k}$ when $s_\alpha > 0$ ($s_\alpha < 0$).
In either case, the presence of the externally imposed gradient leads to a positive contribution to the growth of the ring radius in~\eref{eq:alint1}. 
Hence, beyond a threshold radius satisfying $R_{\rm c}/\ln(R_{\rm c}/a) = |s_\alpha|/(2|\bi{k}|)$, this expanding force overcomes the curvature-induced contraction and the ring expands indefinitely.

When $\nabla\theta_{\rm c}$ results from other defect loops located far away from $\bi{X}_\alpha$, it is given at the leading order in the far-field approximation by a superposition of dipoles, namely,
\begin{equation} \label{eq:dipole}
    \nabla\theta_{\rm c} \simeq \sum_{\beta \ne \alpha}  \frac{3 (\hat\vr_{\alpha \beta}\cdot \bi{m}_\beta)\hat\vr_{\alpha \beta}-\bi{m}_\beta}{r_{\alpha \beta}^3},
\end{equation}
where $\vr_{\alpha \beta} = \bi{X}_\alpha - \bi{X}_\beta$.
In this case too, the loop is globally immobile but only contracts and rotates according to
\begin{eqnarray}
\label{eq:int1}
    \dot{R}_\alpha & = -\frac{1}{R_\alpha} - \sum_{\beta\ne \alpha}\frac{s_\beta}{s_\alpha}\frac{\pi R_\beta^2}{ \ln(R_\alpha/a)}\frac{3 (\hat\vr_{\alpha \beta}\cdot \hat\bi{b}_\alpha)(\hat\vr_{\alpha \beta}\cdot \hat\bi{b}_\beta)-\hat\bi{b}_\alpha\cdot\hat\bi{b}_\beta}{r_{\alpha \beta}^3} ,\\
\label{eq:int2}
    \dot{\hat{\bi{b}}}_\alpha &= \sum_{\beta\ne \alpha}\frac{s_\beta}{s_\alpha}
    \frac{\pi R_\beta^2}{R_\alpha \ln(R_\alpha/a)} \frac{\hat{\bi{b}}_\alpha\times\left[\hat{\bi{b}}_\alpha\times\left(3 (\hat\vr_{\alpha \beta}\cdot \hat\bi{b}_\beta)\hat\vr_{\alpha \beta}-\hat\bi{b}_\beta\right)\right]}{r_{\alpha \beta}^3}.
\end{eqnarray}
In contrast to straight disclination lines (Eq.~\eref{eq_straight_deflines}), the interaction force between loops scales as the inverse of the cube of their separation. 
In particular, the alignment dynamics of disclination loops resembles that of magnetic dipoles,
such that loops with equal and opposite charge will anti-align and align, respectively.
Equation~\eref{eq:int1} further shows that the contraction of the ring is also affected by the presence of nearby loops.
However, given {two} loops $\alpha$ and $\beta$, the requirement for the interaction contribution to compete with the curvature term is $R_\alpha R_\beta^2 \simeq r_{\alpha \beta}^3\ln(R_\alpha/a)$,
which clearly breaks the far field assumption $r_{\alpha \beta} \gg R_\alpha, R_\beta$.
Hence, in most cases distant loops will only rotate and self-annihilate over a finite time.

\section{Discussion}

The reduced particle-field description of the large-scale dynamics of the field theory~\eref{eq_EL} requires a considerably reduced number of degrees of freedom.
Our previous work~\cite{romano2023} was devoted to determining the perturbation of the order parameter introduced by a collection of moving defects. Here, we have focused on the other facet of the problem, namely, determining how defects are set into motion by the order parameter landscape.
As we have demonstrated, for a large class of systems the defect dynamics takes a universal form (Eqs.~\eref{var5} and~\eref{lineq}) since it is largely determined by the large-scale features of the theory, and should thus be qualitatively insensitive to the microscopic details of the system of interest.
One important requirement for~(\ref{var5},\ref{lineq}) to hold is that the underlying dynamics is described by a translationally invariant free energy functional.
We note, however, that coupling the order parameter to a smoothly varying external field should not introduce major difficulties to the derivation presented in Sec.~\ref{sec_exact_var}. 
As a number of nonequilibrium (active) field theories take the form of an equilibrium-like order parameter dynamics coupled to an external flow~\cite{ActiveGelReview,DoostmohammadiActNem}, density~\cite{chateDADAM}, or chemical~\cite{Golestanian2019phoretic,WeberRepPhys2019} field,
some progress could be achieved in these cases via the approach outlined in Sec.~\ref{sec_exact_var}. 
When the order parameter evolution does not satisfy this structure, however, one cannot {\it a priori} exclude that the functional form of the defect equations of motion depends on the details of the microscopic scale physics~\cite{MahaultPRL2018,ChardacPRX2021}.

In Sec.~\ref{sec_def_GL}, we have shown how the nonlinearity of the defect friction leads to a generic dependency on the length scale $\lambda(t)$ (Eq.~\eref{isomotion}).
Hence, a quantitative characterization of the many-body defect dynamics seems out of reach, since it would in principle require to evaluate this scale, which depends on the full history of the defect motion~\cite{romano2023}.
As this task is typically unfeasible---except for relatively simple configurations---we have focused on slowly moving defects and showed how memory effects disappear in this case.
Within this approximation, $\lambda(t)$ can be replaced by any other relevant macroscopic scale of the problem, such that the equation of motion for the defect will be fully determined.

Taking the slow defect approximation moreover allowed us to address the case of defects evolving in a medium exhibiting elastic anisotropy in Sec.~\ref{sec_def_NL}.
Working perturbatively in the anisotropy parameter $\alpha$, we were able to quantify its linear order contribution to the bending of defect trajectories due to the energetic cost of mismatching.
Interestingly, we also discovered that the $+\frac{1}{2}$ nematic defect mobility becomes anisotropic, as it depends on the relative orientation of the comet shape of the defect relatively to the background orientation field.
\Correction{We moreover argue in~\ref{mobilandsymm} that this property may extend to $-1$ defects in polar systems.}
As backflows also substantially affect the dynamics of defects ~\cite{TothPRL2002,Sven2003PRL},
observing the effect of elastic anisotropy experimentally may only be achieved when hydrodynamic effects are negligible, such as in Langmuir monolayers~\cite{Brugues2008PRL}.

The low mobility expansion also made the derivation of the closed form of the equation of motion for disclination lines in Sec.~\ref{sec_discl_lines} more straightforward.
The Peach-Koehler force on the r.h.s.\ of Eq.~\eref{slowandstraexp} takes a similar form as that proposed for disclination lines in three-dimensional nematics~\cite{Long2021SM}, suggesting an interesting connection between the two settings.
Our results moreover indicate that the proportionality of the friction coefficient and the surface tension observed for domain walls extends to diclination lines.
Contrary to point defects in two dimensions, we have shown that line defects freely rotating in three dimensions will always annihilate regardless of their charge.
Furthermore, the study of closed loops done in Sec.~\ref{sec_disc_loops} reveals that they align their moment similarly to magnetic dipoles,
while in the absence of an externally imposed phase gradient they always self-annihilate due to surface tension.
 
Further extensions of the formalism presented in this work can include taking into account the roles of backflow~\cite{TothPRL2002}, fluctuations~\cite{MuznyPRL1992}, or curved geometry~\cite{BowickAdvPhys2009}.
\Correction{
Another interesting extension of the method we propose would be to study the motion of dislocations~\cite{LUBARDA20191550}
by using a field theoretical treatment of elasticity~\cite{HalperinPRL1978}, 
or more sophisticated descriptions~\cite{skogvoll2023unified,paget2023complex}.}
While some of these developments may be technically challenging, they should not entail any additional conceptual difficulty.

\appendix

\addtocontents{toc}{\addtolength{\cftsecnumwidth}{5em}} 
\addtocontents{toc}{\addtolength{\cftsubsecnumwidth}{4em}} 

\section{Symmetry of the mobility matrix from the defect symmetries}
\label{mobilandsymm}
We claim in the main text that the mobility of a defect is generally isotropic. 
We present here an argument supporting this claim, based on the symmetries of the defects.

An isolated $s$-charged defect is $n$-fold symmetric, 
with $n=|s-1|$ and $n=2|s-1|$ for polar and nematic order parameters, respectively.
When $n\neq 1$, the symmetry is nontrivial so that the order parameter field resulting from the presence of the defect must be invariant under $2\pi/n$-rotations. 
\Correction{
In practice, this property implies that the 
defect mobility matrix $\bmu$ satisfies:
\begin{equation}
\label{rotin}
    \bi{R}_n^{T}\bmu\bi{R}_n=\bmu,
\end{equation}
where $\bi{R}_n$ corresponds to the rotation matrix with angle $2\pi/n$.
For $n > 2$ and noting that $\bmu$ is a symmetric matrix, this relation is satisfied only if $\bmu = \mu \bi{I}$, with $\bi{I}$ the identity.
On the other hand, for $n = 1$ or $2$~\eref{rotin} is always satisfied.
We thus conclude that the mobilities of defects with charge $s = 1$ in polar and $s=-\frac{1}{2}$ in nematic systems must be isotropic.
Inversely, the mobilities of $+\frac{1}{2}$-charged nematic and $-1$-charged polar defects may, in principle, be anisotropic.
Focusing on nematic systems and defining $\hat{\p}$ as the polarization of the $+\frac{1}{2}$ defect (see Sec.~\ref{sec_mob_alpha2}), we express its mobility as
\begin{equation}
    \label{positivemob}
    \mu_{ij}\left(\frac{r_0}{a}\right)=\mu_{\|}\left(\frac{r_0}{a}\right)\hat p_i\hat p_j + \mu_{\perp}\left(\frac{r_0}{a}\right)(\delta_{ij} - \hat p_i\hat p_j),
\end{equation}
where, as in the main text, $r_0$ and $a$ denote the matching and core length scales.
It is also worth noting that the polarization $\hat{\p}$ is defined respectively to the orientation of the background nematic field (see Eq.~\eref{eq_nu_plus}), and is therefore not an independent degree of freedom of the dynamics.
As discussed in Sec.~\ref{sec_mob_alpha2}, $\mu_{\|} \ne \mu_{\perp}$ generally occurs in two-dimensional nematics with elastic anisotropy. 

In polar systems, the orientation of the $2$-fold symmetric $-1$ defect can be defined with an appropriate director, which will lead to a mobility matrix of the form of~\eref{positivemob}. 
In principle, a similar 
line of thought can be applied to other 1 and 2-fold symmetric defects. 
However, these are usually unstable due to their charge (like the $s=2$ polar defect), and can thus be observed only with specific choices of boundary conditions \cite{Kralj2017}.}

\section{Integration of the stress tensor for the isotropic case}
\label{isotropicforce}

In this appendix, we derive the dominant term on the r.h.s.\ of Eq.~\eref{var5} for the linear Ginzburg-Landau theory
in the limit $r_0 \to 0$.
As sketched in the main text, we split the near-field orientation gradient into a discontinuous and continuous parts: $\nabla\theta=\nabla\theta_{\rm d}+\nabla\theta_{\rm c}$.
Substituting this expression into $\bi{T}_{\rm bulk}$, and keeping only the non-vanishing contribution in the limit $r_0 \to 0$, we have
\begin{equation}
\fl \qquad \oint_{{\cal C}_{r_0}}\rmd S_j T^{ij}_{\rm bulk}
= \oint_{{\cal C}_{r_0}}\rmd S_j \left(\partial_i\theta_{\rm d}\partial_j\theta_{\rm d}-\frac{1}{2}\delta_{ij}|\nabla\theta_{\rm d}|^2 +
\partial_i\theta_{\rm d}\partial_j\theta_{\rm c}-\delta_{ij}\nabla\theta_{\rm d}\cdot\nabla\theta_{\rm c}\right)
\end{equation}
Using the expression of the discontinuous part given in~\eref{discpart}, 
the mixed term part becomes
\begin{eqnarray}
   \int_{{\cal C}_{r_0}}\rmd S_j \left(\partial_i\theta_{\rm d}\partial_j\theta_{\rm c}-\delta_{ij}\nabla\theta_{\rm d}\cdot\nabla\theta_{\rm c}\right) & =
   \epsilon_{il}\epsilon_{mj}\int_{{\cal C}_{r_0}}\rmd S_j \, \partial_m\theta_{\rm d}\partial_l\theta_{\rm c} \nonumber \\
   & = \epsilon_{il}\partial_l\theta_{\rm c}(\q,t) \int_{{\cal C}_{r_0}}dS_j \,\epsilon_{mj}\partial_m\theta_{\rm d} \nonumber \\
   & = 2\pi s\epsilon_{il}\partial_l\theta_{\rm c}(\q,t),     
\end{eqnarray}
where in the second equality we have used the identity $\epsilon_{il}\epsilon_{mj}=\delta_{im}\delta_{jl}-\delta_{ij}\delta_{lm}$, 
while in the next one was obtained by using the continuity of $\nabla\theta_{\rm c}$ to bring it outside of the integral for vanishing $r_0$.
For the remaining contribution, using \eref{discpart} we find after straightforward calculations 
\begin{equation}
\oint_{{\cal C}_{r_0}}\rmd S_j \left(\partial_i\theta_{\rm d}\partial_j\theta_{\rm d}-\frac{1}{2}\delta_{ij}|\nabla\theta_{\rm d}|^2\right)=\pi s^2 v_{i}(t)\ln\lt(\frac{e^{\frac{1}{2}}\lambda(t)}{r_0}\rt)
\end{equation}

Replacing these expressions in~\eref{var5}, we thus have
\begin{equation}
\label{almosteom}
    \zeta_{ij}\left(\frac{r_0}{a}\right)v_j(t) = -\pi s^2 v_{i}(t) \ln\lt(\frac{e^{\frac{1}{2}}\lambda(t)}{r_0}\rt) - 2\pi s\epsilon_{il}\partial_l\theta_{\rm c}(\q).
\end{equation}
Equation.~\eref{eq_iso_r0} is finally obtained after rearranging the terms of this equation.

\section{Calculation of the pairwise force between defects in an anisotropic medium}
\label{anisforce}

In this appendix, we calculate the integral on the r.h.s.\ of Eq.~\eref{eq_logaexp} for a pair of defects evolving in a system described by the bulk free energy~\eref{simplefrank}
perturbatively in the anisotropy parameter $\alpha$.
Without loss of generality, we consider two oppositely charged defects at positions $\q_\pm = \pm q \hat{\x}$.
The static equation of motion for the orientation field $\theta$ deriving from~\eref{simplefrank} reads
\begin{equation} \label{eq_app_steady_ani}
    \Delta \theta +\sqrt{2}\alpha\left[\Q_{ij}(\theta)\partial^2_{ij}\theta + \tilde{\Q}_{ij}(\theta)(\partial_i\theta)(\partial_j\theta)\right] = 0,
\end{equation}
where 
\begin{equation*}
\tilde{\bQ}(\theta) \equiv \frac{1}{2}\frac{\rmd\bQ}{\rmd\theta} = \frac{1}{\sqrt{2}}\left(\begin{array}{cc}
-\sin 2\theta & \cos2\theta \\
\cos2\theta & \sin2\theta  \\
\end{array} \right).
\end{equation*}

We assume $\alpha$ to be small, so that we write the solution of Eq.~\eref{eq_app_steady_ani} at linear order as $\theta(\x)\simeq\theta_0(\x)+\alpha \theta_1(\x)$,
where $\theta_0$ solves the isotropic ($\alpha = 0$) problem. Namely,
\begin{equation}
    \theta_0(\x,q)=\theta_{\rm d}(\x,q)+\theta_{\rm c},
\end{equation}
where $\theta_{\rm c}$ denotes the continuous part of the solution which has to be constant in the static case, while $\theta_{\rm d}(\x,q) = \frac{1}{2}\arg(\x - q\hat{\x})-\frac{1}{2}\arg(\x + q\hat{\x})$.
In turn, the first order perturbation $\theta_1$ is solution of 
\begin{equation}
\label{thetapert}
\Delta \theta_1 = -\sqrt{2}\left[\Q_{ij}(\theta_0)\partial^2_{ij}\theta_0 + \tilde{\Q}_{ij}(\theta_0)(\partial_i\theta_0)(\partial_j\theta_0)\right].
\end{equation}
Equation~\eref{thetapert} can in principle be solved by inverting the Laplacian.

Here, we however adopt an alternative approach. We first note that, since the bulk theory is conformal, the solution of~\eref{thetapert} can generally be written as
$\theta_1(\x, q, \theta_{\rm c})=\vartheta_1\left(\y,\theta_{\rm c}\right)$,
where $\y = \x/q$ and $\vartheta_1\left(\y,\theta_{\rm c}\right)$ is the solution of~\eref{thetapert} for two defects at positions $\pm \hat{\x}$.
From the following property of the $\bQ$ and $\tilde{\bQ}$ tensors,
\begin{eqnarray*}
    \bQ(\theta_{\rm d}+\theta_{\rm c})=\cos(2\theta_{\rm c})\bQ(\theta_{\rm d})+\sin(2\theta_{\rm c})\tilde{\bQ}(\theta_{\rm d}),\\
    \tilde{\bQ}(\theta_{\rm d}+\theta_{\rm c})=\cos(2\theta_{\rm c})\tilde{\bQ}(\theta_{\rm d})-\sin(2\theta_{\rm c})\bQ(\theta_{\rm d}),
\end{eqnarray*}
we moreover write $\vartheta_1$ as
\begin{equation}
\label{fact}
\vartheta_1\left(\y,\theta_{\rm c}\right)=\cos(2\theta_{\rm c})\vartheta_{\rm cs}(\y)+\sin(2\theta_{\rm c})\vartheta_{\rm sn}(\y),
\end{equation}
where  $\vartheta_{\rm cs}(\y)$ and $\vartheta_{\rm sn}(\y)$ are solutions of the equations
\begin{eqnarray}
\label{eq_app_thetacs}
\fl  \;  \Delta \vartheta_{\rm cs}(\y) & =- \sqrt{2}\left[\Q_{ij}(\theta_{\rm d}(\y,1))\partial^2_{ij}\theta_{\rm d}(\y,1) + \tilde{\Q}_{ij}(\theta_{\rm d}(\y,1))(\partial_i\theta_{\rm d}(\y,1))(\partial_j\theta_{\rm d}(\y,1)) \right],\\
\label{eq_app_thetasn}
\fl  \;  \Delta \vartheta_{\rm sn}(\y) & = -\sqrt{2}\left[\tilde{\Q}_{ij}(\theta_{\rm d}(\y,1))\partial^2_{ij}\theta_{\rm d}(\y,1) - \Q_{ij}(\theta_{\rm d}(\y,1))(\partial_i\theta_{\rm d}(\y,1))(\partial_j\theta_{\rm d}(\y,1)) \right].
\end{eqnarray}
These two equations can be formally solved using the Green's function of the Laplacian. 
As will appear clear below, the solutions are importantly independent of $q$ and $\theta_{\rm c}$.

We are now able to evaluate the integral of the stress tensor of Eq.~\eref{eq_logaexp}.
Since this integral is independent of the integration contour, we choose it to be the straight line along $y$ passing by $x = 0$.
The r.h.s.\ of Eq.~\eref{eq_logaexp} can then be written as
\begin{equation} \label{eq_app_T_contour}
-\int_{{\cal C}}\rmd S_j T_{{\rm sb},ij} = -\int^{+\infty}_{-\infty}\rmd y \, \hat{x}_j T_{{\rm sb},ij} = -\int^{+\infty}_{-\infty}\rmd y \, T_{{\rm sb},i1}.
\end{equation}
We moreover also expand $\bi{T}_{{\rm sb}}$ as $\bi{T}_{{\rm sb}} = \bi{T}_{0} + \alpha \bi{T}_{1}$
so that $\bi{T}_{0}$ is the stress tensor of the unperturbed theory and $\bi{T}_{1}$ the corresponding linear order perturbation.
The integral for $\bi{T}_{0}$ is straightforward, and gives the usual Coulomb interaction.
To calculate the perturbation, we first expand $\bi{T}_{1} = \frac{1}{q}[\bi{T}_{\rm cs}\cos(2\theta_{\rm c}) +  \bi{T}_{\rm sn}\sin(2\theta_{\rm c})]$,
which after some straightforward algebra leads to
\begin{eqnarray}
    T_{{\rm cs},ij} = & (\partial_i \vartheta_{\rm cs})(\partial_j\theta_{\rm d}) + (\partial_j \vartheta_{\rm cs})(\partial_i\theta_{\rm d}) - \delta_{ij}(\partial_k \vartheta_{\rm cs}) (\partial_k\theta_{\rm d}) \nonumber \\
    & + \sqrt{2}(\partial_i\theta_{\rm d})\Q_{jk}(\theta_{\rm d})(\partial_k\theta_{\rm d}) - \frac{1}{\sqrt{2}}\delta_{ij}(\partial_k\theta_{\rm d}) \Q_{kl}(\theta_{\rm d})(\partial_l\theta_{\rm d}) ,\\
    T_{{\rm sn},ij} = & (\partial_i \vartheta_{\rm sn})(\partial_j\theta_{\rm d}) + (\partial_j \vartheta_{\rm sn})(\partial_i\theta_{\rm d}) - \delta_{ij}(\partial_k \vartheta_{\rm sn}) (\partial_k\theta_{\rm d}) \nonumber \\
    & + \sqrt{2}(\partial_i\theta_{\rm d})\tilde{\Q}_{jk}(\theta_{\rm d})(\partial_k\theta_{\rm d}) - \frac{1}{\sqrt{2}}\delta_{ij}(\partial_k\theta_{\rm d}) \tilde{\Q}_{kl}(\theta_{\rm d})(\partial_l\theta_{\rm d}) ,
\end{eqnarray}
where we kept the dependencies of the fields in $\y$ implicit. Replacing these expressions into~\eref{eq_app_T_contour} and $\vartheta_{cs}$ and $\vartheta_{sn}$ by the solutions of Eqs.~(\ref{eq_app_thetacs},\ref{eq_app_thetasn}), 
we are able to express the force between defects in terms of intricate integrals.
However, these integrals are independent of the two parameters of the problem: $q$ and $\theta_{\rm c}$, 
such that they essentially amount to numerical coefficients in the final equation.
Evaluating them numerically, we obtain
\begin{equation}
    \int^{+\infty}_{-\infty}\rmd y \, T_{{\rm cs},i1} = 0 , \qquad \int^{+\infty}_{-\infty}\rmd y \, T_{{\rm sn},i1} \approx -1.0472\delta_{i2} \approx -\frac{\pi}{3}\delta_{i2}.
\end{equation}
Going back to the lab frame, we finally recover the expression given in Eq.~\eref{anismotion} for an arbitrary orientation of the defect pair.

\section{Details on numerical simulations}

\label{app_numerics}

To perform the numerical simulations whose results are presented in Secs.~\ref{sec_sc_two_defects} and~\ref{sec_misalignment_forces}, 
we mapped the two-dimensional order parameter $\bphi$ onto the complex number $f = \rho e^{i n \theta}$ with $n = 1$ or $2$ for polar or nematic order, respectively.
Introducing the complex variable $z = x+i y$ and the corresponding derivative $\partial \equiv \partial_x + i \partial_y$, 
the free energy associated with the dynamics becomes:
\begin{equation} \label{eq_app_Fc}
\fl \qquad\quad  \calF_{\rm c}= \int \rmd z\,\rmd z^* \, \left[ \frac{1}{2}(\partial f^*)(\partial^* f) + \frac{\alpha}{2}\delta_{n,2}{\rm Re}[f^* (\partial f)(\partial f^*)]+ \chi^2 \left(1-|f|^2\right)^2 \right],
\end{equation}
where stars denote complex conjugate,
while the dynamics of $f$ is simply obtained via
\begin{equation} \label{eq_app_eom_f}
 \fl \qquad\quad \partial_t f = -2\,\frac{\delta \calF_{\rm c}}{\delta f^*} = 4\chi^2(1 - |f|^2)f + \Delta f + \alpha \delta_{n,2}\left[ {\rm Re}(f^* \partial^2)f + \frac{1}{2}(\partial^* f)^2 \right].
\end{equation}

Equation~\eref{eq_app_eom_f} was solved in a periodic box via a pseudo-spectral method and an explicit fourth order Runge-Kutta scheme.
In the isotropic case (Fig.~\ref{GL_trajecotries}), simulations were performed in a system of size $4096 \times 4096$. For an initial separation $q(0)=64$ we set spatial and temporal resolutions to $\rmd x = \frac{1}{4}$ and $\rmd t = 0.004$, while for $q(0)=32$ we have $\rmd x = \frac{1}{8}$ and $\rmd t = 0.001$.
The simulations at finite anisotropy (Fig.~\ref{bigcurves}) were  all performed in a system of size $2048 \times 2048$ with $\rmd x = \frac{1}{4}$ and $\rmd t = 0.004$.

The coefficient of the quartic potential in~\eref{eq_app_Fc} was set to $\chi = \sqrt{10}$ in all the simulations.
All simulations were initialized with the following profile
\begin{equation}
\label{incond}
    f(z,t=0) = e^{i\theta(z)}, \qquad \theta(z) = \arg(z+q_0) - \arg(z-q_0),
\end{equation} 
which corresponds to a pair of defects with charges $s = \pm n^{-1}$ at positions $\pm q_0$ on the complex plane.
To reduce the importance of finite size effects, in the anisotropic case the initial distance between the defects was taken to be $2|q_0| = 64$, {\it i.e.}  much less than the linear dimensions of the whole system.

The results presented in Sec.~\ref{sec_sc_two_defects} (Fig.~\ref{GL_trajecotries}) were obtained from simulations of~\eref{eq_app_eom_f} with $n = 1$, 
which thus reduces to the classical Ginzburg-Landau theory.
In turn, simulations corresponding to Sec.~\ref{sec_sc_two_defects} (Fig.~\ref{bigcurves}) were performed with $n=2$ and varying the anisotropy parameter $\alpha$.
As in both cases~\eref{incond} is not a solution of Eq.~\eref{eq_app_eom_f} on the torus, we initially let the dynamics evolve over a simulation time 
corresponding to roughly $10\%$ of the time need by the defect pair to annihilate before starting the data acquisition.
Defect tracking was performed by computing the circuitation of the phase of $f$ around four neighbouring boxes of the numerical grid.

\section*{References}
\bibliography{bibliography.bib}

\providecommand{\newblock}{}
\begin{thebibliography}{10}
\expandafter\ifx\csname url\endcsname\relax
  \def\url#1{{\tt #1}}\fi
\expandafter\ifx\csname urlprefix\endcsname\relax\def\urlprefix{URL }\fi
\providecommand{\eprint}[2][]{\url{#2}}

\bibitem{Nelson2002defects}
Nelson D~R 2002 {\em Defects and geometry in condensed matter physics\/}
  (Cambridge University Press)

\bibitem{HadzibabicNature2006}
Hadzibabic Z, Kr{\"u}ger P, Cheneau M, Battelier B and Dalibard J 2006 {\em
  Nature\/} {\bf 441} 1118--1121
  \urlprefix\url{https://doi.org/10.1038/nature04851}

\bibitem{Hindmarsh_1995}
Hindmarsh M~B and Kibble T~W~B 1995 {\em Rep. Prog. Phys.\/} {\bf 58} 477
  \urlprefix\url{https://dx.doi.org/10.1088/0034-4885/58/5/001}

\bibitem{Bray2002review}
Bray A~J 2002 {\em Adv. Phys.\/} {\bf 51} 481--587
  \urlprefix\url{https://doi.org/10.1080/00018730110117433}

\bibitem{Kosterlitz1973JPC}
Kosterlitz J~M and Thouless D~J 1973 {\em Journal of Physics C: Solid State
  Physics\/} {\bf 6} 1181--1203
  \urlprefix\url{https://doi.org/10.1088/0022-3719/6/7/010}

\bibitem{Abrikosov2004RMP}
Abrikosov A~A 2004 {\em Rev. Mod. Phys.\/} {\bf 76}(3) 975--979
  \urlprefix\url{https://link.aps.org/doi/10.1103/RevModPhys.76.975}

\bibitem{uchida2010}
Uchida N and Golestanian R 2010 {\em Phys. Rev. Lett.\/} {\bf 104}(17) 178103
  \urlprefix\url{https://link.aps.org/doi/10.1103/PhysRevLett.104.178103}

\bibitem{Wensink2012}
Wensink H~H, Dunkel J, Heidenreich S, Drescher K, Goldstein R~E, L\"{o}wen H
  and Yeomans J~M 2012 {\em Proceedings of the National Academy of Sciences\/}
  {\bf 109} 14308--14313
  \urlprefix\url{https://doi.org/10.1073/pnas.1202032109}

\bibitem{SanchezNature2012}
Sanchez T, Chen D~T~N, DeCamp S~J, Heymann M and Dogic Z 2012 {\em Nature\/}
  {\bf 491} 431--434 \urlprefix\url{https://doi.org/10.1038/nature11591}

\bibitem{Giomi2013}
Giomi L, Bowick M~J, Ma X and Marchetti M~C 2013 {\em Phys. Rev. Lett.\/} {\bf
  110}(22) 228101
  \urlprefix\url{https://link.aps.org/doi/10.1103/PhysRevLett.110.228101}

\bibitem{Thampi2013}
Thampi S~P, Golestanian R and Yeomans J~M 2013 {\em Phys. Rev. Lett.\/} {\bf
  111}(11) 118101
  \urlprefix\url{https://link.aps.org/doi/10.1103/PhysRevLett.111.118101}

\bibitem{Kawaguchi2017nature}
Kawaguchi K, Kageyama R and Sano M 2017 {\em Nature\/} {\bf 545} 327--331
  \urlprefix\url{https://doi.org/10.1038/nature22321}

\bibitem{Saw2017Nature}
Saw T~B, Doostmohammadi A, Nier V, Kocgozlu L, Thampi S, Toyama Y, Marcq P, Lim
  C~T, Yeomans J~M and Ladoux B 2017 {\em Nature\/} {\bf 544} 212--216
  \urlprefix\url{https://doi.org/10.1038/nature21718}

\bibitem{MaroudasSacks2020}
Maroudas-Sacks Y, Garion L, Shani-Zerbib L, Livshits A, Braun E and Keren K
  2020 {\em Nature Physics\/} {\bf 17} 251--259
  \urlprefix\url{https://doi.org/10.1038/s41567-020-01083-1}

\bibitem{CopenhagenNatPhys2021}
Copenhagen K, Alert R, Wingreen N~S and Shaevitz J~W 2021 {\em Nat. Phys.\/}
  {\bf 17} 211--215 \urlprefix\url{https://doi.org/10.1038/s41567-020-01056-4}

\bibitem{PismenBook}
Pismen L~M 1999 {\em {Vortices in nonlinear fields: From liquid crystals to
  superfluids. From nonequilibrium patterns to cosmic strings}\/} (Oxford
  University Press)

\bibitem{Dafermos1970}
Dafermos C~M 1970 {\em Q. J. Mech. Appl. Math.\/} {\bf 23} 49--64 ISSN
  0033-5614 \urlprefix\url{https://doi.org/10.1093/qjmam/23.2.49}

\bibitem{IMURA1973403}
Imura H and Okano K 1973 {\em Phys. Lett. A\/} {\bf 42} 403--404 ISSN 0375-9601
  \urlprefix\url{https://www.sciencedirect.com/science/article/pii/0375960173907287}

\bibitem{Eshelby1980PhilMagA}
Eshelby J~D 1980 {\em Philos. Mag. A\/} {\bf 42} 359--367
  \urlprefix\url{https://doi.org/10.1080/01418618008239363}

\bibitem{AmbegaokarPRB1980}
Ambegaokar V, Halperin B~I, Nelson D~R and Siggia E~D 1980 {\em Phys. Rev. B\/}
  {\bf 21}(5) 1806--1826
  \urlprefix\url{https://link.aps.org/doi/10.1103/PhysRevB.21.1806}

\bibitem{Dubois-violette}
Dubois-violette E, Guazzelli E and Prost J 1983 {\em Philos. Mag. A\/} {\bf 48}
  727--747 \urlprefix\url{https://doi.org/10.1080/01418618308236540}

\bibitem{Kawasaki1983linedef}
Kawasaki K 1983 {\em Physica A\/} {\bf 119} 17--40 ISSN 0378-4371
  \urlprefix\url{https://www.sciencedirect.com/science/article/pii/0378437183901437}

\bibitem{Kawasaki1984Progr}
Kawasaki K 1984 {\em Prog. Theor. Phys. Supp.\/} {\bf 79} 161--190 ISSN
  0375-9687 \urlprefix\url{https://doi.org/10.1143/PTPS.79.161}

\bibitem{KAWASAKI1984319}
Kawasaki K 1984 {\em Ann. Phys. (N. Y.)\/} {\bf 154} 319--355 ISSN 0003-4916
  \urlprefix\url{https://www.sciencedirect.com/science/article/pii/0003491684901544}

\bibitem{BODENSCHATZ1988PhysD}
Bodenschatz E, Pesch W and Kramer L 1988 {\em Physica D\/} {\bf 32} 135--145
  ISSN 0167-2789
  \urlprefix\url{https://www.sciencedirect.com/science/article/pii/0167278988900905}

\bibitem{NEU1990PhysD}
Neu J~C 1990 {\em Physica D\/} {\bf 43} 385--406 ISSN 0167-2789
  \urlprefix\url{https://www.sciencedirect.com/science/article/pii/016727899090143D}

\bibitem{Pismen1990PRA}
Pismen L~M and Rodriguez J~D 1990 {\em Phys. Rev. A\/} {\bf 42}(4) 2471--2474
  \urlprefix\url{https://link.aps.org/doi/10.1103/PhysRevA.42.2471}

\bibitem{Rubinstein1991}
Rubinstein J 1991 {\em Quart. Appl. Math.\/} {\bf 49} 1--9
  \urlprefix\url{https://doi.org/10.1090/qam/1096227}

\bibitem{RodriguezPRA1991}
Rodriguez J~D, Pismen L~M and Sirovich L 1991 {\em Phys. Rev. A\/} {\bf 44}(12)
  7980--7984 \urlprefix\url{https://link.aps.org/doi/10.1103/PhysRevA.44.7980}

\bibitem{Pismen1991PhysicaD}
Pismen L and Rubinstein J 1991 {\em Physica D\/} {\bf 47} 353--360 ISSN
  0167-2789
  \urlprefix\url{https://www.sciencedirect.com/science/article/pii/0167278991900358}

\bibitem{DennistonPRB1996}
Denniston C 1996 {\em Phys. Rev. B\/} {\bf 54}(9) 6272--6275
  \urlprefix\url{https://link.aps.org/doi/10.1103/PhysRevB.54.6272}

\bibitem{Pleiner1988PRA}
Pleiner H 1988 {\em Phys. Rev. A\/} {\bf 37}(10) 3986--3992
  \urlprefix\url{https://link.aps.org/doi/10.1103/PhysRevA.37.3986}

\bibitem{Semenov_1999EPL}
Semenov A~N 1999 {\em EPL\/} {\bf 46} 631
  \urlprefix\url{https://dx.doi.org/10.1209/epl/i1999-00312-y}

\bibitem{NajafiEPJB2003}
Najafi A and Golestanian R 2003 {\em Eur. Phys. J. B\/} {\bf 34} 99--103
  \urlprefix\url{https://doi.org/10.1140/epjb/e2003-00200-x}

\bibitem{Radzihovsky2015PRL}
Radzihovsky L 2015 {\em Phys. Rev. Lett.\/} {\bf 115}(24) 247801
  \urlprefix\url{https://link.aps.org/doi/10.1103/PhysRevLett.115.247801}

\bibitem{PismenPRE2013}
Pismen L~M 2013 {\em Phys. Rev. E\/} {\bf 88}(5) 050502
  \urlprefix\url{https://link.aps.org/doi/10.1103/PhysRevE.88.050502}

\bibitem{TangSM2019}
Tang X and Selinger J~V 2019 {\em Soft Matter\/} {\bf 15}(4) 587--601
  \urlprefix\url{http://dx.doi.org/10.1039/C8SM01901K}

\bibitem{CortesePRE2018}
Cortese D, Eggers J and Liverpool T~B 2018 {\em Phys. Rev. E\/} {\bf 97}(2)
  022704 \urlprefix\url{https://link.aps.org/doi/10.1103/PhysRevE.97.022704}

\bibitem{ShankarPRL2018}
Shankar S, Ramaswamy S, Marchetti M~C and Bowick M~J 2018 {\em Phys. Rev.
  Lett.\/} {\bf 121}(10) 108002
  \urlprefix\url{https://link.aps.org/doi/10.1103/PhysRevLett.121.108002}

\bibitem{Vafadefects2020}
Vafa F, Bowick M~J, Marchetti M~C and Shraiman B~I 2020 Multi-defect dynamics
  in active nematics \urlprefix\url{https://arxiv.org/abs/2007.02947}

\bibitem{ZhangPRE2020}
Zhang Y~H, Deserno M and Tu Z~C 2020 {\em Phys. Rev. E\/} {\bf 102}(1) 012607
  \urlprefix\url{https://link.aps.org/doi/10.1103/PhysRevE.102.012607}

\bibitem{AnghelutaNJP2021}
Angheluta L, Chen Z, Marchetti M~C and Bowick M~J 2021 {\em New J. Phys.\/}
  {\bf 23} 033009 \urlprefix\url{https://doi.org/10.1088/1367-2630/abe8a8}

\bibitem{VafaSoftMatt2022}
Vafa F 2022 {\em Soft Matter\/} {\bf 18}(42) 8087--8097
  \urlprefix\url{http://dx.doi.org/10.1039/D2SM00830K}

\bibitem{Gartland2002ws}
Gartland Jr E~C, Sonnet A and Virga E~G 2002 {\em Continuum Mech. Thermodyn.\/}
  {\bf 14} 307--319 \urlprefix\url{https://doi.org/10.1007/s00161-002-0099-8}

\bibitem{romano2023}
Romano J, Mahault B and Golestanian R 2023 {\em J. Stat. Mech.: Theory Exp.\/}
  {\bf 2023} 083211 \urlprefix\url{https://dx.doi.org/10.1088/1742-5468/aceb57}

\bibitem{Sven2003PRL}
Sven\ifmmode~\check{s}\else \v{s}\fi{}ek D and \ifmmode~\check{Z}\else
  \v{Z}\fi{}umer S 2003 {\em Phys. Rev. Lett.\/} {\bf 90}(15) 155501
  \urlprefix\url{https://link.aps.org/doi/10.1103/PhysRevLett.90.155501}

\bibitem{Brugues2008PRL}
Brugu\'es J, Ign\'es-Mullol J, Casademunt J and Sagu\'es F 2008 {\em Phys. Rev.
  Lett.\/} {\bf 100}(3) 037801
  \urlprefix\url{https://link.aps.org/doi/10.1103/PhysRevLett.100.037801}

\bibitem{MissaouiPRR2020}
Missaoui A, Harth K, Salamon P and Stannarius R 2020 {\em Phys. Rev.
  Research\/} {\bf 2}(1) 013080
  \urlprefix\url{https://link.aps.org/doi/10.1103/PhysRevResearch.2.013080}

\bibitem{Review_LC_Harth2020}
Harth K and Stannarius R 2020 {\em Front. Phys.\/} {\bf 8} ISSN 2296-424X
  \urlprefix\url{https://www.frontiersin.org/articles/10.3389/fphy.2020.00112}

\bibitem{ALLEN1979}
Allen S~M and Cahn J~W 1979 {\em Acta Metall.\/} {\bf 27} 1085--1095 ISSN
  0001-6160
  \urlprefix\url{https://www.sciencedirect.com/science/article/pii/0001616079901962}

\bibitem{HH1977}
Hohenberg P~C and Halperin B~I 1977 {\em Rev. Mod. Phys.\/} {\bf 49}(3)
  435--479 \urlprefix\url{https://link.aps.org/doi/10.1103/RevModPhys.49.435}

\bibitem{toupin1960stress}
Toupin R 1960 {\em Archive for Rational Mechanics and Analysis\/} {\bf 5}
  440--452

\bibitem{susskind2017special}
Susskind L and Friedman A 2017 {\em Special relativity and classical field
  theory\/} (Penguin UK)

\bibitem{LUBARDA20191550}
Lubarda V~A 2019 {\em J. Mater. Res. Technol.\/} {\bf 8} 1550--1565 ISSN
  2238-7854
  \urlprefix\url{https://www.sciencedirect.com/science/article/pii/S2238785418304551}

\bibitem{Ryskin1991PRL}
Ryskin G and Kremenetsky M 1991 {\em Phys. Rev. Lett.\/} {\bf 67}(12)
  1574--1577
  \urlprefix\url{https://link.aps.org/doi/10.1103/PhysRevLett.67.1574}

\bibitem{Dzubiella2003PRL}
Dzubiella J, L\"owen H and Likos C~N 2003 {\em Phys. Rev. Lett.\/} {\bf 91}(24)
  248301 \urlprefix\url{https://link.aps.org/doi/10.1103/PhysRevLett.91.248301}

\bibitem{soto2014PRL}
Soto R and Golestanian R 2014 {\em Phys. Rev. Lett.\/} {\bf 112}(6) 068301
  \urlprefix\url{https://link.aps.org/doi/10.1103/PhysRevLett.112.068301}

\bibitem{Gupta2022PRE}
Gupta R~K, Kant R, Soni H, Sood A~K and Ramaswamy S 2022 {\em Phys. Rev. E\/}
  {\bf 105}(6) 064602
  \urlprefix\url{https://link.aps.org/doi/10.1103/PhysRevE.105.064602}

\bibitem{deGennes_Prost}
de~Gennes P and Prost J 1993 {\em The Physics of Liquid Crystals\/}
  International Series of Monographs on Physics (Clarendon Press) ISBN
  9780198517856 \urlprefix\url{https://books.google.de/books?id=0Nw-dzWz5agC}

\bibitem{LandauToE}
Landau L~D, Pitaevskii L~P, Lifshitz E~M and Kosevich A~M 1986 {\em Theory of
  Elasticity\/} 3rd ed (Butterworth-Heinemann) ISBN 075062633X
  \urlprefix\url{http://www.amazon.com/Theory-Elasticity-Third-Theoretical-Physics/dp/075062633X/ref=sr_1_16?ie=UTF8&s=books&qid=1280929419&sr=8-16}

\bibitem{Aranson2022RMP}
Aranson I~S and Kramer L 2002 {\em Rev. Mod. Phys.\/} {\bf 74}(1) 99--143
  \urlprefix\url{https://link.aps.org/doi/10.1103/RevModPhys.74.99}

\bibitem{Fetter2001BEC}
Fetter A~L and Svidzinsky A~A 2001 {\em J. Phys. Condens. Matter\/} {\bf 13}
  R135 \urlprefix\url{https://dx.doi.org/10.1088/0953-8984/13/12/201}

\bibitem{Wu2023NatCom}
Wu Z~W, Chen Y, Wang W~H, Kob W and Xu L 2023 {\em Nat. Commun.\/} {\bf 14}
  2955 \urlprefix\url{https://doi.org/10.1038/s41467-023-38547-w}

\bibitem{ActiveGelReview}
Prost J, J{\"u}licher F and Joanny J~F 2015 {\em Nat. Phys.\/} {\bf 11}
  111--117 \urlprefix\url{https://doi.org/10.1038/nphys3224}

\bibitem{DoostmohammadiActNem}
Doostmohammadi A, Ign{\'e}s-Mullol J, Yeomans J~M and Sagu{\'e}s F 2018 {\em
  Nat. Commun.\/} {\bf 9} 3246 ISSN 2041-1723
  \urlprefix\url{https://doi.org/10.1038/s41467-018-05666-8}

\bibitem{chateDADAM}
Chat{\'e} H and Mahault B 2022 {Dry, Aligning, Dilute, Active Matter: A
  Synthetic and Self-contained Overview} {\em {Active Matter and Nonequilibrium
  Statistical Physics: Lecture Notes of the Les Houches Summer School: Volume
  112, September 2018}\/} (Oxford University Press) ISBN 9780192858313
  \urlprefix\url{https://doi.org/10.1093/oso/9780192858313.003.0001}

\bibitem{Golestanian2019phoretic}
Golestanian R 2022 {Phoretic Active Matter} {\em {Active Matter and
  Nonequilibrium Statistical Physics: Lecture Notes of the Les Houches Summer
  School: Volume 112, September 2018}\/} (Oxford University Press) ISBN
  9780192858313
  \urlprefix\url{https://doi.org/10.1093/oso/9780192858313.003.0008}

\bibitem{WeberRepPhys2019}
Weber C~A, Zwicker D, J{\"u}licher F and Lee C~F 2019 {\em Rep. Prog. Phys.\/}
  {\bf 82} 064601 \urlprefix\url{https://dx.doi.org/10.1088/1361-6633/ab052b}

\bibitem{MahaultPRL2018}
Mahault B, Jiang X~c, Bertin E, Ma Y~q, Patelli A, Shi X~q and Chat\'e H 2018
  {\em Phys. Rev. Lett.\/} {\bf 120}(25) 258002
  \urlprefix\url{https://link.aps.org/doi/10.1103/PhysRevLett.120.258002}

\bibitem{ChardacPRX2021}
Chardac A, Hoffmann L~A, Poupart Y, Giomi L and Bartolo D 2021 {\em Phys. Rev.
  X\/} {\bf 11}(3) 031069
  \urlprefix\url{https://link.aps.org/doi/10.1103/PhysRevX.11.031069}

\bibitem{TothPRL2002}
T\'oth G, Denniston C and Yeomans J~M 2002 {\em Phys. Rev. Lett.\/} {\bf
  88}(10) 105504
  \urlprefix\url{https://link.aps.org/doi/10.1103/PhysRevLett.88.105504}

\bibitem{Long2021SM}
Long C, Tang X, Selinger R~L~B and Selinger J~V 2021 {\em Soft Matter\/} {\bf
  17}(8) 2265--2278 \urlprefix\url{http://dx.doi.org/10.1039/D0SM01899F}

\bibitem{MuznyPRL1992}
Muzny C~D and Clark N~A 1992 {\em Phys. Rev. Lett.\/} {\bf 68}(6) 804--807
  \urlprefix\url{https://link.aps.org/doi/10.1103/PhysRevLett.68.804}

\bibitem{BowickAdvPhys2009}
Bowick M~J and Giomi L 2009 {\em Adv. Phys.\/} {\bf 58} 449--563
  \urlprefix\url{https://doi.org/10.1080/00018730903043166}

\bibitem{HalperinPRL1978}
Halperin B~I and Nelson D~R 1978 {\em Phys. Rev. Lett.\/} {\bf 41}(2) 121--124
  \urlprefix\url{https://link.aps.org/doi/10.1103/PhysRevLett.41.121}

\bibitem{skogvoll2023unified}
Skogvoll V, R{\o}nning J, Salvalaglio M and Angheluta L 2023 {\em npj
  Computational Materials\/} {\bf 9} 122 ISSN 2057-3960
  \urlprefix\url{https://doi.org/10.1038/s41524-023-01077-6}

\bibitem{paget2023complex}
Paget J, Mazza M~G, Archer A~J and Shendruk T~N 2023 {\em Nature
  Communications\/} {\bf 14} 1048

\bibitem{Kralj2017}
Kralj S, Murray B~S and Rosenblatt C 2017 {\em Phys. Rev. E\/} {\bf 95}(4)
  042702 \urlprefix\url{https://link.aps.org/doi/10.1103/PhysRevE.95.042702}

\end{thebibliography}

\end{document}